\documentclass[a4paper,sort&compress,oneside,preprint,12pt,number]{elsarticle}

\usepackage{amssymb,amsmath,amsthm,mathtools}
\usepackage{subfigure,overpic}
\usepackage{hyperref}
\newcommand{\vareps}{\varepsilon}

\usepackage[utf8]{inputenc}
\usepackage[english]{babel}
\usepackage{amsmath}
\usepackage{amsfonts}
\usepackage{amssymb}
\usepackage{graphicx}
\usepackage{lmodern}
\usepackage{kpfonts}
\usepackage[left=2cm,right=2cm,top=2cm,bottom=2cm]{geometry}

\def\XXint#1#2#3{{\setbox0=\hbox{$#1{#2#3}{\int}$}
     \vcenter{\hbox{$#2#3$}}\kern-.5\wd0}}

\newcommand{\dd}{\partial}

\newcommand{\la}{\lambda}
\newcommand{\non}{\nonumber}
\newcommand{\eps}{\epsilon}
\newcommand{\beqa}{\begin{eqnarray}}
\newcommand{\eeqa}{\end{eqnarray}}
\newcommand{\beqas}{\begin{eqnarray*}}
\newcommand{\eeqas}{\end{eqnarray*}}
\newcommand{\ba}{\begin{align}}
\newcommand{\ea}{\end{align}}
\newcommand{\bas}{\begin{align*}}
\newcommand{\eas}{\end{align*}}
\newcommand{\beq}{\begin{equation}}
\newcommand{\eeq}{\end{equation}}
\newcommand{\re}{\mathrm{Re}}
\newcommand{\im}{\mathrm{Im}}
\newcommand{\ra}{\rightarrow}

\newcommand{\ee}{\mathrm{e}}
\newcommand{\ii}{\mathrm{i}}
\newcommand{\pdhfrac}[2]{\mathchoice{\frac{#1}{#2}}{#1/#2}{#1/#2}{#1/#2}}
\newcommand{\fdd}[2]{\pdhfrac{\mathrm{d}#1}{\mathrm{d}#2}}

\newcommand{\pd}[2]{\pdhfrac{{\partial}#1}{{\partial}#2}}

\renewcommand{\d}[1]{\mathrm{d}#1}
\DeclareMathOperator{\sech}{sech}
\DeclareMathOperator{\Ai}{Ai}
\DeclareMathOperator{\Bi}{Bi}
\DeclareSymbolFont{matha}{OML}{txmi}{m}{it}
\DeclareMathSymbol{\varv}{\mathord}{matha}{118}

\journal{Nonlinearity}

\begin{document}

\begin{frontmatter}



\title{Self-similar blow-up solutions in the %
generalized Korteweg-de Vries equation: Spectral analysis,
normal form and asymptotics}



\author{S.~Jon~Chapman}
\affiliation{organization={Mathematical Institute, %
            University of Oxford, AWB, ROQ},       
            addressline={Woodstock Road},
            city={Oxford},
            state={OX2 6GG},
            country={UK}}

\author{M.~Kavousanakis}
\affiliation{organization={School of Chemical Engineering, %
             National Technical University of Athens},
            city={Athens},
            postcode={15780},
            country={GR}}

\author{E.G.~Charalampidis}
\affiliation{organization={Mathematics Department, %
             California Polytechnic State University},
            city={San Luis Obispo},
            postcode={93407-0403},
            state={CA},
            country={USA}}

\author{I.G.~Kevrekidis}
\affiliation{organization={Department of Chemical and %
                           Biomolecular Engineering \& \\
                           Department of Applied Mathematics %
                           and Statistics, Johns Hopkins University},
            city={Baltimore},
            postcode={21218},
            state={MD},
            country={USA}}

\author{P.G.~Kevrekidis}
\affiliation{organization={Department of Mathematics and Statistics, %
                          University of Massachusetts},
            city={Amherst},
            postcode={1003-4515},
            state={MA},
            country={USA}}

\begin{abstract}
In the present work we revisit the problem of
the
generalized Korteweg-de Vries equation parametrically, as a function
of the relevant nonlinearity exponent, to examine the emergence of
blow-up solutions, as traveling waveforms lose their stability past
a critical point of the relevant parameter $p$, here at $p=5$.
We provide a {\it normal form} of the associated collapse dynamics
and illustrate how this captures the collapsing branch bifurcating
from the unstable traveling branch. We also systematically
characterize
the linearization spectrum of not only the traveling states, but
importantly
of the emergent collapsing waveforms in the so-called co-exploding
frame where these waveforms are identified as stationary states.
This spectrum, in addition to
two positive real eigenvalues which are shown to be associated
with the symmetries of translation and scaling invariance of the
original
(non-exploding) frame features complex patterns of negative
eigenvalues that we also fully characterize. We show that the
phenomenology of the latter is significantly affected by the boundary
conditions and is far more complicated than in the corresponding
symmetric Laplacian case of the nonlinear Schr{\"o}dinger problem
that has recently been explored.
In addition, we explore the dynamics of the
unstable solitary waves for $p>5$
in the co-exploding
frame.
\end{abstract}




\end{frontmatter}



\section{Introduction}\label{Intro}\label{sec_1}

This study centers around the investigation of the spectral stability of solutions to the
generalized Korteweg-de Vries equation, denoted as gKdV hereafter:
\begin{align}
\label{eq:genKdV}
\frac{\partial u}{\partial t} = -\frac{\partial^3 u}{\partial x^3} %
- \frac{\partial u^p}{\partial x},
\end{align}
where the nonlinearity exponent, $p\geq 2$, governs the behavior of solutions that exhibit self-similar blow-up in finite time.
The gKdV equation is a prototypical dispersive nonlinear partial
differential
equation (PDE) which possesses solitary waveforms due to the interplay of
dispersion and
nonlinearity~\cite{ablowitz_clarkson_1991,whitham_1974}.
In addition to its central relevance to the description of water
waves~\cite{water,dauxois}, it also emerges through
the continuum limit approximation~\cite{zabusky_kruskal_1965} of the well-known
Fermi-Pasta-Ulam-Tsingou (FPUT) lattice model~\cite{FPUT,gallavotti_2008}
in which the $p$ specifies the nonlinear exponent governing the FPUT's  force law ~\cite{zabusky}.
Specifically, for $p=2$ and $p=3$ the gKdV corresponds to the
classical KdV~\cite{KdV_original} and the modified KdV (mKdV) models, respectively.
KdV and mKdV are both integrable ~\cite{hirota_2004,ablowitz_clarkson_1991,mKdV_Olver}
in the sense that they have an infinite number of conserved quantities
and can be described as compatibility conditions of suitable
Lax pairs~\cite{ablowitz_clarkson_1991,water,dauxois}.
The
practical applications of KdV and mKdV span a broad range of fields, including
shallow water dynamics, optical fibers, plasma physics, ion-acoustic solitons, and
electric circuits, among others%
~\cite{whitham_1974,water,dauxois,remoissenet_1999,osborne_2010,nakamura_1985}.

While the KdV and mKdV equations are integrable, the gKdV equation for other values of
the nonlinearity exponent, $p$, lacks integrability, to the best of
our current understanding.
However, when considering smooth
functions that decay suitably fast as $|x|\rightarrow\infty$,
the gKdV possesses three fundamental conserved
quantities:
\begin{subequations}
\begin{align}
\label{consv_1}
E_{1}&\coloneqq\|u\|_{1}=\int_{-\infty}^{\infty}u\,dx,
\\%
\label{consv_2}
E_{2}&\coloneqq\|u\|^{2}_{2}=\int_{-\infty}^{\infty}u^{2}\,dx,
\\
\label{consv_3}
E_{3}&\coloneqq \frac{1}{2}\int_{-\infty}^{\infty}%
\left[u_{x}^{2}-\frac{1}{p+1}u^{p+1}\right]\,dx.
\end{align}
\end{subequations}

These quantities correspond to the conservation of mass, momentum and energy,
respectively~\cite{bona_1987,weinmueller_2020}.
The invariance of the gKdV equation under space and time translations
is responsible, through Noether theory~\cite{sulem}, for the second and
third
among these conservation laws, while the fundamental form of the gKdV as
a conservation law leads to the conservation of mass.
However, in addition to these key conserved quantities,
the model also possesses a scale invariance:
\begin{align}
\label{scale_invariance}
u(x,t)\mapsto \lambda^{\frac{2}{p-1}}u(\lambda x,\lambda^{3}t), \quad \lambda\neq 0.
\end{align}
%
It is important to highlight that this transformation preserves the
momentum in the critical case of $p=5$ that will be further discussed below.
This
transformation group also
leads to the notable fact that if $u(x,t)$ is a valid solution,
then $u(-x,-t)$ is also a solution.

Even though the gKdV equation is not known to be integrable for most values of the nonlinearity exponent, $p$ (with the notable exceptions
of $p=2$ and $p=3$), it admits traveling wave (TW) solutions
parametrically as a function of $p\geq 1$.
These TW solutions are characterized by the following analytically available expression:
\begin{align}
\label{eq:soliton}
u(x,t) \coloneqq Q(\xi)= %
\left(\frac{c\left(p+1\right)}{2}\right)^{\frac{1}{p-1}}%
\sech^{\frac{2}{p-1}}{\left(\frac{\sqrt{c}\left(p-1\right)}{2}\xi\right)},
\end{align}
where $\xi\coloneqq x-ct$, and $c$ denotes the wave speed.
In~\cite{bona_1987},
it was shown that the TW solution of Eq.~\eqref{eq:soliton} with $c=1$ is
linearly unstable for $p\geq 5$.
Furthermore, it becomes evident that blow-up in finite time is possible when
$p>5$, corresponding to the super-critical case~\cite{merle_2001}.
If $p=5$, i.e.,
the critical case, solutions $u$ to the gKdV exhibit global existence and temporal boundedness,
granted that the initial data $u_{0}$ satisfy the condition $\|u_{0}\|_{2}<\|Q\|_{2}$~\cite{martel_merle_2002}.
This criterion is based on a Gagliardo-Nirenberg
inequality argument~\cite{weinstein_1983}.
Based on the same argument, it
can be shown that solutions also globally exist and remain bounded for the subcritical
case, $p<5$.

Numerous efforts have been dedicated to numerically investigating blow-up solutions within the context of the gKdV equation; we briefly mention a few here.
In the work of ~\cite{bona_dougalis_1995}, a comprehensive approach to studying the instability of solitary waves leading to blow-up in a similarity form was introduced.
This involved employing high-order numerical schemes in both spatial and time dimensions, supplemented by mesh
adaptivity.
The study centered on tracking the growth rate of various
$L^{s}$-norms of the
solution.
The numerical computations carried out in
~\cite{dix_mckinney_1998} delved deeper into this phenomenon, and discovered
how self-similar blow-up takes place for rapidly decaying solutions at
$|x|\rightarrow\infty$.
Notably, ~\cite{dix_mckinney_1998} argued that this
blow-up phenomenon is intricately linked to the instability of TWs, and that the self-similar solutions
inherit the stability that the TW solutions have relinquished.
In parallel with the computation of self-similar profiles in~\cite{dix_mckinney_1998}, the work presented
in~\cite{koch_2015} approached the self-similar blow-up problem within the gKdV framework as a bifurcation problem, and constructed an approximate invariant manifold that
encompasses both TW and self-similar blow-up solutions.
In~\cite{klein_peter_2015},
a numerical study of the stability of TWs was carried out by performing
direct numerical simulations using exponential time differencing
methods ~\cite{klein_2008}. This was also the first study that
performed computations of the fully dynamically rescaled form of the PDE.
A significant finding in ~\cite{klein_peter_2015} was that TWs
are unstable against being radiated away and blowing-up, thus numerically identifying
the blow-up mechanism discussed in detail
in~\cite{martel_merle_raphael_part_I,martel_merle_raphael_part_II}.
However, additional intriguing features were also identified such as
the
emergence of a dispersive shock wave (prior to collapse) in the small
dispersion
regime.

More recently, the research presented in~\cite{weinmueller_2020} has
delved into the supercritical regime, $p>5$, resolving the structure
of self-similar blow-up solutions,
upon computing these solutions systematically as stationary ones.
Their work involved the implementation of a dynamic rescaling
technique
using collocation and finite difference methods~\cite{bvpsuite1} and was
complemented by a detailed analysis of the solutions' asymptotics (in space).
Indeed, and alongside the sech-like shape of the self-similar
profile in its core, the authors found that the solution features a slow algebraic
decay to the left of the peak and a rapid (exponentially dominated)
decay to the right~\cite{weinmueller_2020}.

In the present paper, we expand upon the above works, and not only focus on
the construction of self-similar blow-up solutions within the gKdV framework, but also
study their spectral stability, as well as obtain the normal form associated with
the bifurcation that the TW solution~\eqref{eq:soliton} undergoes at
the critical point of $p=5$, in order to elucidate the onset of
collapse. More specifically, we continue the long-term program of
systematically characterizing the emergence of self-similarly
collapsing
waveforms as a bifurcation problem through the instability of regular
solitary waveforms. This was initiated in the work of~\cite{siettos},
which
built on the earlier seminal works of~\cite{Pap,Pap1,Pap3} for the
widely considered case of the nonlinear Schr{\"o}dinger (NLS) model. The
bifurcation
structure of the problem was computed numerically and the spectral
analysis of the emergent collapsing solutions in the self-similar
frame was given. This was subsequently further elucidated in the work
of~\cite{jon1} which obtained systematically, leveraging
asymptotics beyond all orders, the normal form dynamics
of the NLS model near the bifurcation point giving rise to the
emergence
of self-similarly collapsing waveforms.  Lastly, in the context of
NLS, our recent work of~\cite{chapman_2022} led to a detailed
understanding of both the point and the continuous spectrum
of the stationary states in the self-similar frame, attributing the
putative instabilities therein to the breaking (within this
co-exploding
frame) of original symmetries of translational and scaling invariance; see
also~\cite{bernoff}.

Here, we accomplish both of the following goals for the generalized
KdV model. We extract the associated normal form and showcase its
pitchfork-like structure similarly to NLS, although we illustrate
that, differently from the NLS case, exponentially small terms do not
arise in the present gKdV setting. In the normal form identified,
linear terms ---emerging with an absolute value---
and quadratic ones (in the blow-up rate $G$) are found to
dominate
the right hand side and the associated phenomenology near the
bifurcation
point. The symmetry of the normal form reflects the existence
of solutions with both $G$ and $-G\neq0$ (i.e., the ``pitchfork-like'' feature of the normal form), yet the linear and
quadratic terms are also reminiscent of a transcritical normal
form.

At the same time, we provide a detailed analysis of the
linearization spectrum not only around the simpler solitonic
waveforms, but also around the self-similarly collapsing states
emerging past the critical bifurcation point of $p=5$. The latter
states are found to feature a particularly complicated spectrum
(especially as concerns the continuous spectrum), driven
by the third derivative operator of the gKdV problem and the
associated boundary conditions that we use for the collapsing
waveforms. That turns out to be fundamentally different once again
from the  simpler case of the NLS problem, although the two
cases share the same type of apparent (yet dynamically innocuous) instabilities,
stemming from the invariance with respect to translation and the
invariance with respect to scaling of the original frame (as viewed
in the co-exploding frame where the collapsing solution is
stationary).

Our presentation is structured as follows. In the next
section, we provide the overarching scaling formulation, connecting
the traveling waves of the original frame and the self-similar
solutions upon dynamic rescaling. We set up the associated stationary
problems and boundary conditions thereof and also illustrate how
to formulate the corresponding linearization problem. Then, the
core of our findings is presented in section 3, which starts with the
simpler (and well-known) spectral analysis of the traveling waveforms
and then proceeds to extend considerations to the self-similar
structures. Both the point and the continuous spectrum of the latter
is examined and, finally, the normal form of the associated collapse
dynamics is elucidated. Finally, in section 4, we summarize our
findings
and present our conclusions, as well as a number of directions for
further
study.

\section{Mathematical Setup}\label{sec_2}

We now set the stage for our forthcoming discussions within this work.
Our approach follows the dynamic renormalization technique described in ~\cite{chapman_2022} (and references therein).
This approach allows us to rewrite the problem in the
co-exploding frame through the ansatz:
\begin{align}
\label{eq:scaling}
u(x,t)\coloneqq A(\tau) w \left( \xi, \tau \right), %
\quad \xi\coloneqq \frac{x}{B(\tau)}+K(\tau), %
\quad \tau\coloneqq \tau(t).
\end{align}
Here, $A$ and $B$ represent the solution's amplitude and width, and alongside the function $K$, all depend on the renormalized time, $\tau$.
Upon inserting Eq.~\eqref{eq:scaling} into Eq.~\eqref{eq:genKdV} and employing the chain rule,
we arrive at:
{\small
\begin{align}
\label{eq:gKdVSSF_v0}
\left(\frac{1}{A}\frac{\partial A}{\partial \tau}w%
+\left(K-\xi\right)\frac{1}{A}\frac{\partial A}{\partial \tau}%
\frac{\partial w}{\partial \xi}%
+\frac{\partial K}{\partial \tau}\frac{\partial w}{\partial \xi}
+\frac{\partial w}{\partial \tau}\right)%
\frac{\partial \tau }{\partial t}=-\frac{1}{B^{3}}%
\left(\frac{\partial^{3}w}{\partial \xi^{3}}+A^{p-1}B^{2}\frac{\partial w^{p}}{\partial \xi}\right)
\end{align}
}
\noindent after dividing both sides by $A$ (where $A\neq 0$).
We balance the terms on the
right-hand-side (RHS) of Eq.~\eqref{eq:gKdVSSF_v0} by imposing the condition,
$A^{p-1}B^{2}=1$, or equivalently:
\begin{align}
\label{eq:consscaling}
A=B^{-2/(p-1)},
\end{align}
which leads to:
\begin{align}
\label{eq:A_func}
\frac{1}{A}\frac{\partial A}{\partial \tau}=%
-\frac{2}{p-1}\frac{1}{B}\frac{\partial B}{\partial \tau}=-\frac{2G}{p-1} , \quad
G=G(\tau)\coloneqq \frac{1}{B}\frac{\partial B}{\partial \tau},
\end{align}
where $G$ represents the rate of change of the solution's width, $B$.
Additionally, if we demand time independence of the effective
dynamics in the renormalized frame, it is relevant to use:
\begin{align}
\label{eq:tau_func}
\frac{\partial \tau}{\partial t}=\frac{1}{B^{3}}.
\end{align}
Then, Eq.~\eqref{eq:gKdVSSF_v0} simplifies to the \textit{renormalized gKdV}
equation:
\begin{align}
\label{eq:renKdV_v1}
\frac{\partial w}{\partial \tau} = -\frac{\partial^3 w}{\partial \xi^3}%
-\frac{\partial w^p}{\partial \xi} + G\left(\frac{2w}{p-1}+ %
\xi \frac{\partial w}{\partial \xi} \right)%
-\left(KG+\frac{\partial K}{\partial \tau}\right)\frac{\partial w}{\partial \xi}.
\end{align}
To reduce the dynamics to the case of a TW with speed $c=1$,
we require:
\begin{align}
\label{eq:condition_soliton}
KG+\frac{\partial K}{\partial \tau}=-1.
\end{align}
Notice that this is in line with the formulation of~\cite{koch_2015}
(see Eq.~(6) therein). Importantly, and in order to unify the relevant
formulations in different recent works, we remind the reader that
the transformation discussed therein (in particular,
using $W(y)=G^{-2/(3 p)} w(G^{-{1/3}} y- G^{-1})$) provides a matching
of the problem considered with the setup of~\cite{weinmueller_2020}.

Consequently, Eq.~\eqref{eq:renKdV_v1} transforms, according to Eq.~\eqref{eq:condition_soliton}, into:
\begin{align}
\label{eq:renKdV_v2}
\frac{\partial w}{\partial \tau} = -\frac{\partial^3 w}{\partial \xi^3}%
-\frac{\partial w^p}{\partial \xi} + G\left(\frac{2w}{p-1}+ %
\xi \frac{\partial w}{\partial \xi} \right)%
+\frac{\partial w}{\partial \xi}.
\end{align}
%
%
As per the definition of $G$ in Eq.~\eqref{eq:A_func}, and the monotonic
growth of $\tau$ from Eq.~\eqref{eq:tau_func}, the solution's width, $B$, diminishes, signifying a contraction in its spatial extent - effectively, $B_{\tau}<0$.
%
Accordingly, self-similar blow-up solutions in \textit{forward time} with respect to
Eq.~\eqref{eq:genKdV} correspond to steady-state solutions of Eq.~\eqref{eq:renKdV_v2}
characterized by  $G<0$.
To elaborate, as $t\rightarrow t^{\ast}$ with $t^{\ast}>0$
representing the blow-up time in the original frame, then $\tau\rightarrow \infty$ resulting in $w(\xi,\tau)\mapsto w(\xi)$, i.e.,
reaching a stationary self-similar profile for which $G(\tau)\rightarrow \mathrm{const.}(<0)$.
Furthermore, the transformation $\tau\mapsto -\tau$, $G\mapsto -G$, and
$\xi\mapsto-\xi$ leaves Eq.~\eqref{eq:renKdV_v2} invariant thereby permitting the existence of a self-similar solution that blows up in \textit{backward time}.
In other words, this solution appears to be ``coming back from infinity'' characterized by
$G\rightarrow \mathrm{const}(>0)$.
[The latter will be important towards attributing a pitchfork-like
form to the relevant bifurcation diagram. However, it is important to keep
in mind that the stability of this solution is also reversed in comparison
to the one with $G<0$. Moreover, the linear and quadratic
terms that we will encounter below within the system's reduced
dynamics will also be reminiscent of a
transcritical normal form].

Returning to the stationary problem, when  $|\tau|\rightarrow \infty$,
Eq.~\eqref{eq:renKdV_v2} simplifies into the stationary ordinary differential
equation (ODE):
\begin{align}
\label{eq:renKdV_v3}
-\frac{d^3 w}{d \xi^3}%
-\frac{d w^p}{d \xi} + G\left(\frac{2w}{p-1}+ %
\xi \frac{d w}{d \xi} \right)%
+\frac{d w}{d \xi}=0.
\end{align}
For our numerical computations throughout this work, we consider a symmetric and finite spatial domain $[-L,L]$, where $L\gg 1$.
The bounded nature of the computational domain necessitates the implementation of boundary
conditions (BCs) which in this work are:
\begin{subequations}
\begin{align}
\label{eq:renKdV_BC_1}
\frac{2Gw}{p-1}+\left(G\xi+1\right)\frac{\partial w}{\partial \xi}%
\Bigg|_{|\xi|= L}=0,
\\
\label{eq:renKdV_BC_2}
\frac{\partial^{2}w}{\partial \xi^{2}}\Bigg|_{\xi = L}=0.
\end{align}
\end{subequations}
%
It is important to highlight here that
these boundary conditions
bear similarities to the ones employed  in~\cite{weinmueller_2020},
yet incorporate the role of the blow-up rate $G$ explicitly.
Our subsequent analysis will illustrate that they
accurately encapsulate the asymptotic behaviors of the self-similar profiles that we compute in this work.
It is noteworthy that based on the aforementioned BCs, and considering $G<0$, the solution's slow algebraic decay manifests to the left of its peak, while the more rapid exponential decay transpires on its right side (consistent with the profiles in~\cite{weinmueller_2020}).
If we perform the transformation $G\mapsto -G$ and $\xi\mapsto-\xi$
in Eqs.~\eqref{eq:renKdV_v3} (which retains its invariance)
and~\eqref{eq:renKdV_BC_1}-\eqref{eq:renKdV_BC_2}, these decaying
behaviors interchange, aligning with the numerical findings reported in ~\cite{koch_2015}.
%

It has been reported that $G$ is a dynamic variable, gradually approaching a constant value as  $|\tau|\rightarrow\infty$, which needs to be determined
in a self-consistent manner for a given value of $p$.
To achieve this, following the methodologies outlined in the works of~\cite{siettos,jon1,chapman_2022}
(and references therein), we ``close'' the systems of Eqs.~\eqref{eq:renKdV_v2}
and~\eqref{eq:renKdV_v3} (both subject to the BCs of Eqs.~\eqref{eq:renKdV_BC_1}-%
\eqref{eq:renKdV_BC_1}) by introducing a pinning condition.
This lifts the degeneracy of the one-parameter infinity of available
(self-similarly rescaled) solutions and selects a unique one among
them, while at the same time providing the appropriate value of $G$.
In our subsequent analysis and computations detailed in the forthcoming sections, and without any loss of generality, we enforce a ``pinning condition'' in the form of an internal
BC:
\begin{align}
\label{pinning_cond}
\frac{\partial w}{\partial \xi}\Big|_{\xi=0}=0.
\end{align}
This condition facilitates the unique determination of a self-similar solution characterized by a constant value of $G$.
%
%
%
By constraining the wave's value (or its derivatives) at a specific point, as exemplified in Eq.~\eqref{pinning_cond}, one can effectively compute the pertinent
unique solution which in this case is prescribed to be peaked
(i.e., features a maximum) at the origin.

We conclude this section by outlining the primary framework of the gKdV spectral stability problem.
We investigate the linear stability of a self-similar solution
$w(\xi)$ (characterized by $G\neq 0$), which remains stationary in the self-similar frame.
This study involves considering the ansatz:
\begin{align}
\widetilde{w}(\xi,\tau)=w(\xi)+\delta\,v(\xi)e^{\lambda\tau}, %
\quad \delta\ll 1,
\label{eq:stab_ansatz}
\end{align}
where $(\lambda,v(\xi))$ denotes the eigenvalue-eigenvector pair.
Upon inserting Eq.~\eqref{eq:stab_ansatz} into Eq.~\eqref{eq:renKdV_v2},
we arrive at order $\mathcal{O}(\delta)$ at the eigenvalue problem:
\begin{align}
\label{eq:renKdV_eval_prob}
\lambda\,v=\mathcal{L}v,
\end{align}
where the operator $\mathcal{L}\left(\cdot\right)$ is defined as:
\begin{align}
\label{eq:renKdV_eval_prob_op}
\mathcal{L}\left(\cdot\right)\coloneqq \frac{2G}{p-1}-p\left(p-1\right)w^{p-2}\frac{dw}{d\xi}+%
\left(G\xi+1-pw^{p-1}\right)\frac{d}{d\xi}-\frac{d^{3}}{d\xi^{3}},
\end{align}
representing the linearization operator whose eigenvalues we compute.
Notably,  the stability characteristics of TWs (over $p$) can also be explored using Eq.~\eqref{eq:renKdV_eval_prob_op}, albeit with $G=0$.
For all cases, i.e., $G=0$ or $G\neq 0$, the eigenvalues $\lambda=\lambda_{r}+\ii\lambda_{i}$
of the linear operator presented in~\eqref{eq:renKdV_eval_prob_op} provide information
about the stability of the computed waveforms.
The presence of a non-zero positive real part in an eigenvalue, i.e., $\lambda_{r}>0$, indicates linear instability
in the computed solution.
Conversely, the absence of
eigenvalues with a positive real part suggests (linear) stability.
It is also important to appreciate the ``dual'' character of the
relevant linearization setting, similarly to what was discussed
in the NLS case in~\cite{siettos,chapman_2022}. Namely, the relevant
problem is a Hamiltonian one for $G=0$, featuring the corresponding eigenvalue
symmetries (i.e., for each complex $\lambda$, each of $-\lambda$,
the complex conjugate $\lambda^*$ and $-\lambda^*$ are all eigenvalues).
On the other hand, for $G \neq 0$, the problem is genuinely non-conservative
and only the symmetry of $\lambda \rightarrow \lambda^{*}$ will be present
in the corresponding spectra. This renders the relevant bifurcation
problem especially interesting from a mathematical point of view,
given its mixed Hamiltonian-dissipative type.

In the upcoming section on our numerical findings, we will delve into the challenges posed by the eigenvalue problem in
Eq.~\eqref{eq:renKdV_eval_prob} addressing them as we compute both the
point and the continuous spectrum of the operator $\mathcal{L}$ (as defined in Eq.~\eqref{eq:renKdV_eval_prob_op}).

\section{Computational and Theoretical Results}\label{sec_3}
%
We begin our discussion with a brief presentation of the spectra of
the better-known case of the TWs, i.e., for $G=0$ across various nonlinearity exponents, $p$, as we consider the relevant bifurcation problem.
Subsequently, we delve into a comprehensive analysis of the existence and spectral properties
of self-similar solutions, characterized by $G\neq 0$, and their asymptotic behavior (for $|\xi|\gg 1$) within the co-exploding frame.
We also present the normal form that describes the bifurcating branch of self-similar solutions from the solitary wave branch,
at $p=5$.

\subsection{The TW case and spectra: $G=0$}\label{sec_3_1}
Let us first focus on the soliton case corresponding to $G=0$ in Eqs.~\eqref{eq:renKdV_v3} and~\eqref{eq:renKdV_eval_prob}.
While the analytical expression of the TW solution is available (see, Eq.~\eqref{eq:soliton}),
we numerically compute the solution to Eq.~\eqref{eq:renKdV_v3}
subject to the boundary conditions given by Eqs.~\eqref{eq:renKdV_BC_1}-\eqref{eq:renKdV_BC_2}.
Notice that for the present branch of $G=0$, these
boundary conditions amount to a homogeneous Neumann
boundary condition pair plus the vanishing of the second
derivative on one of the sides (given the need for 3
such conditions).
This approach is taken to ensure that the computation of the TW's spectrum is conducted on the same computational grid and with the same spatial discretization as used for solving the root-finding problem
of Eq.~\eqref{eq:renKdV_v3}.
This strategy eliminates local truncation errors (LTEs) that could arise if we had directly plugged the TW solution into the eigenvalue problem of
Eq.~\eqref{eq:renKdV_eval_prob}, which might have led to potential
perturbations of the eigenvalues.
In this study, we adopt a finite volume discretization approach.
%
For discretizing the computational domain $\xi \in [-L,L]$, we have chosen equidistant nodes with a spacing of $\delta \xi = 0.001$.
We compare the results for different values of $L$ as discussed below.

\begin{figure}[pt]
\centering
\includegraphics[width = 0.49\textwidth]{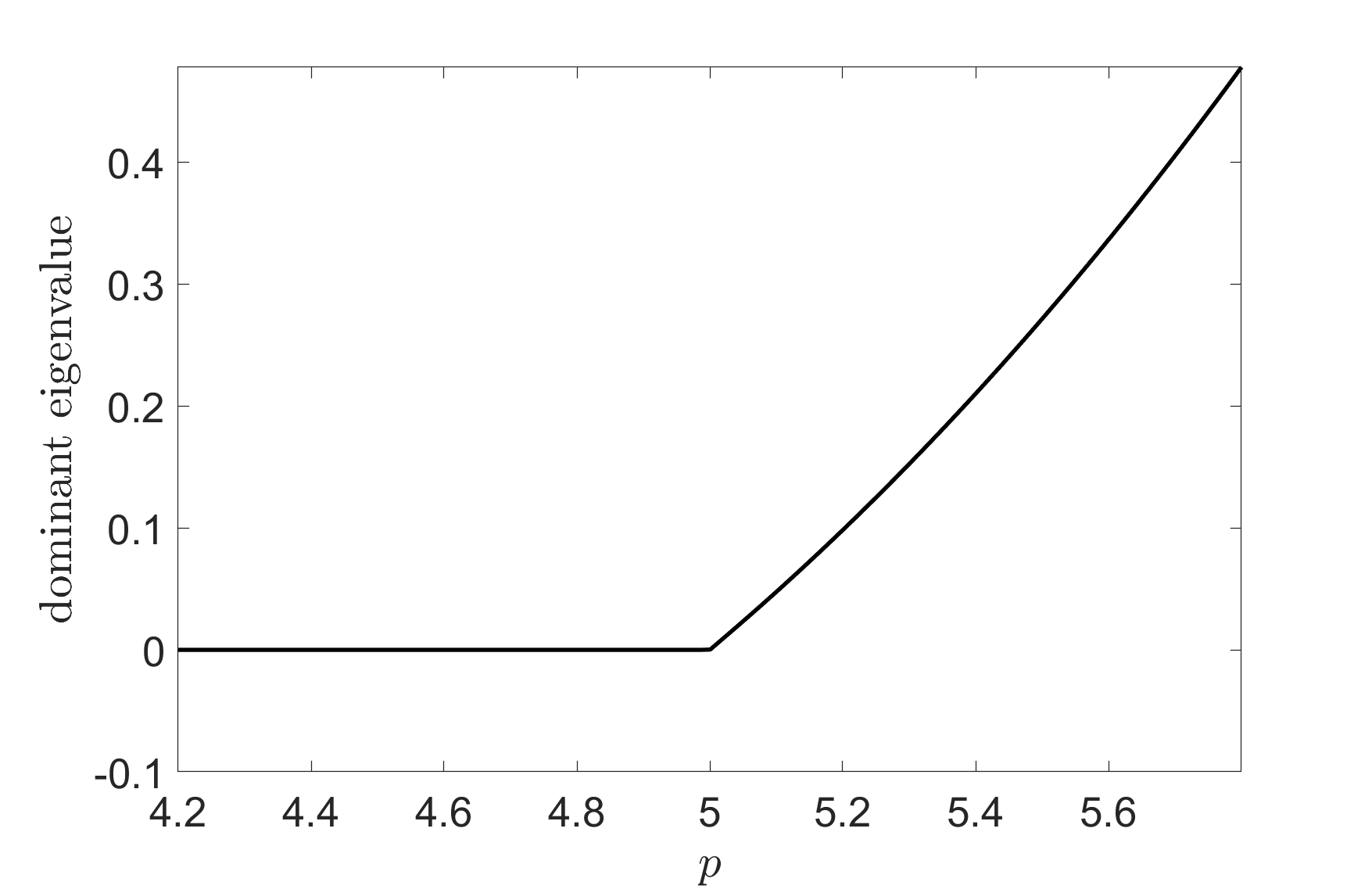}
\includegraphics[width = 0.49\textwidth]{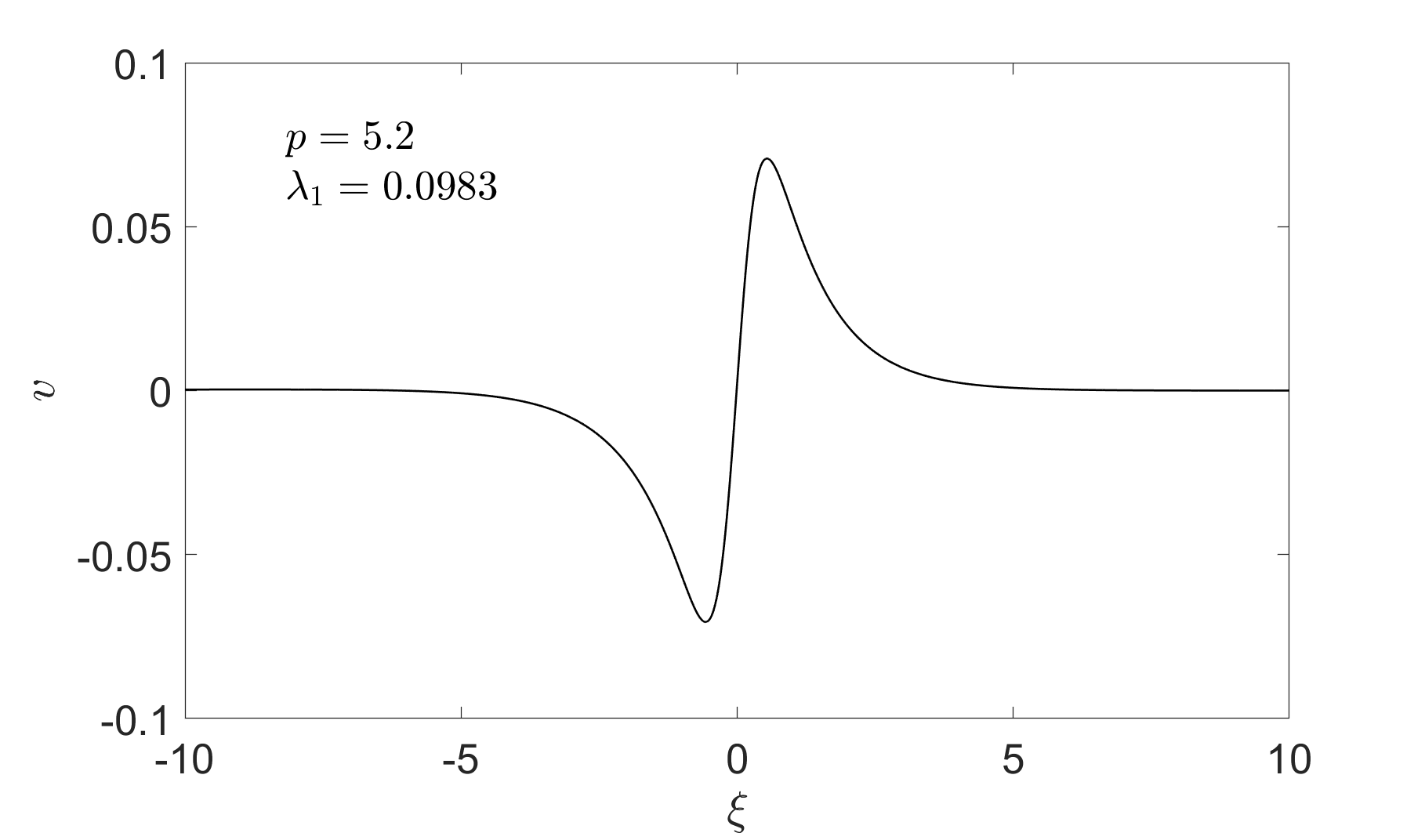}
\caption{
(left panel) The variation of the largest real, i.e., dominant eigenvalue
of the TW solution [cf. Eq.~\eqref{eq:soliton}] over the nonlinearity
exponent $p$. The TW becomes spectrally unstable for $p>5$. The right panel shows the corresponding eigenvector, $v$, for the dominant eigenvalue ($\lambda_1=0.0983$) of the TW solution for $p=5.2$ .
\label{fig:eigenvalue_soliton}}
\end{figure}

We use the TW solution from Eq.~\eqref{eq:soliton} as an initial guess within a Newton algorithm for $p=2$, progressively conducting continuation steps across varying values of $p$ up to $p=5.8$.
At each continuation step, we compute the
spectrum of the operator $\mathcal{L}$ [cf. Eq.~\eqref{eq:renKdV_eval_prob_op}]
associated with the eigenvalue problem defined by Eq.~\eqref{eq:renKdV_eval_prob}
for $G=0$, corresponding to the solitonic branch.
We compute our numerical spectra using MATLAB's \textit{eigs} function.
During this computation, we pay particular attention to the dominant unstable (real) eigenvalue.
In Fig.~\ref{fig:eigenvalue_soliton},
we present the variation of this eigenvalue with respect to $p$, showcasing the results from $p=4.2$ to $5.8$.
The depicted figure clearly illustrates that the TW remains stable until $p=5$, as evidenced by the absence of an unstable  eigenmode with a positive
real part.
However, as we push beyond the threshold of $p=5$, instability sets in, accompanied by the emergence of a positive real eigenmode.
This observation corroborates the findings in~\cite{bona_1987} for the gKdV equation.

\begin{figure}[pt]
\centering
\includegraphics[width = 0.495\textwidth]{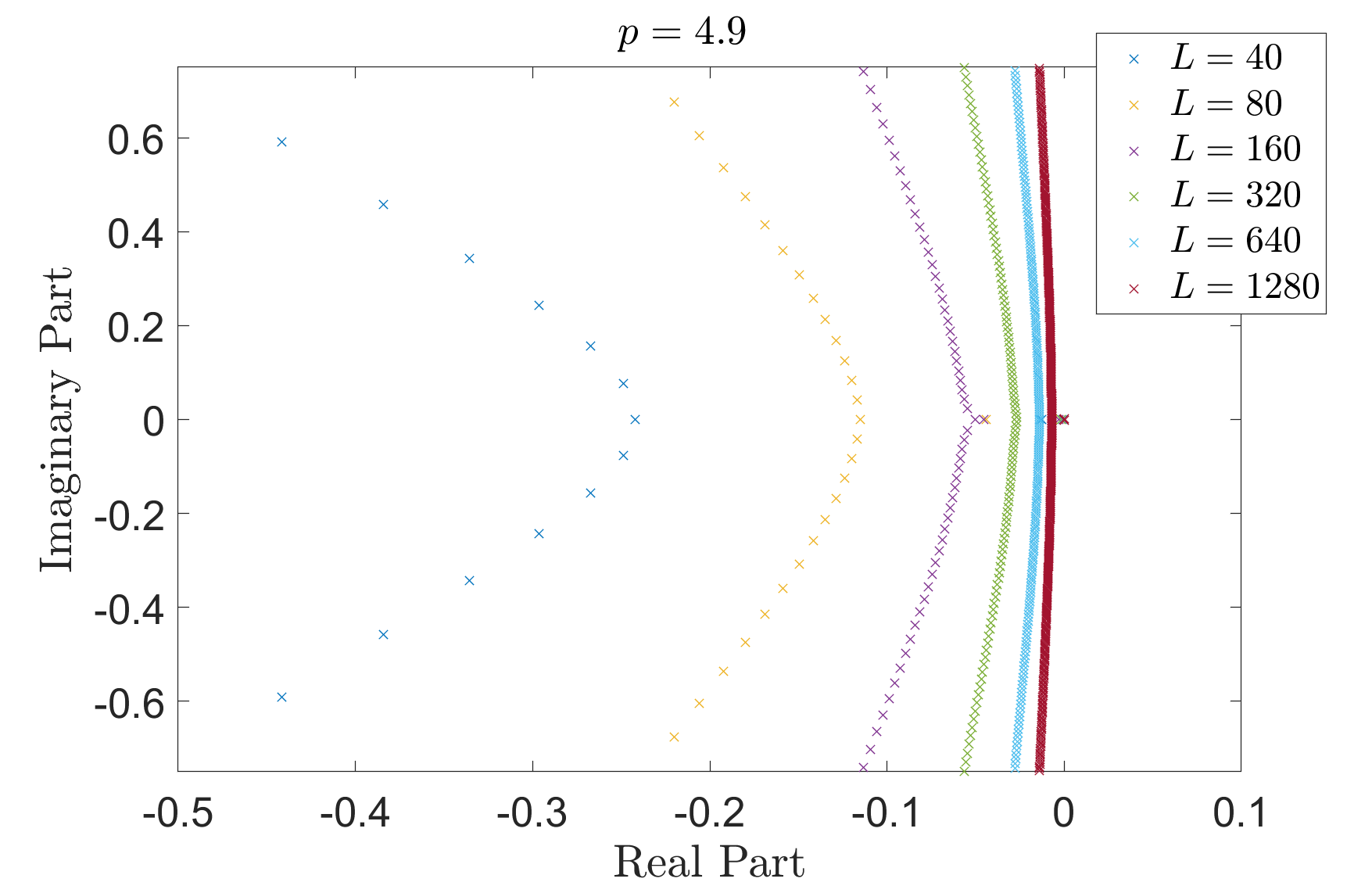}
\includegraphics[width = 0.495\textwidth]{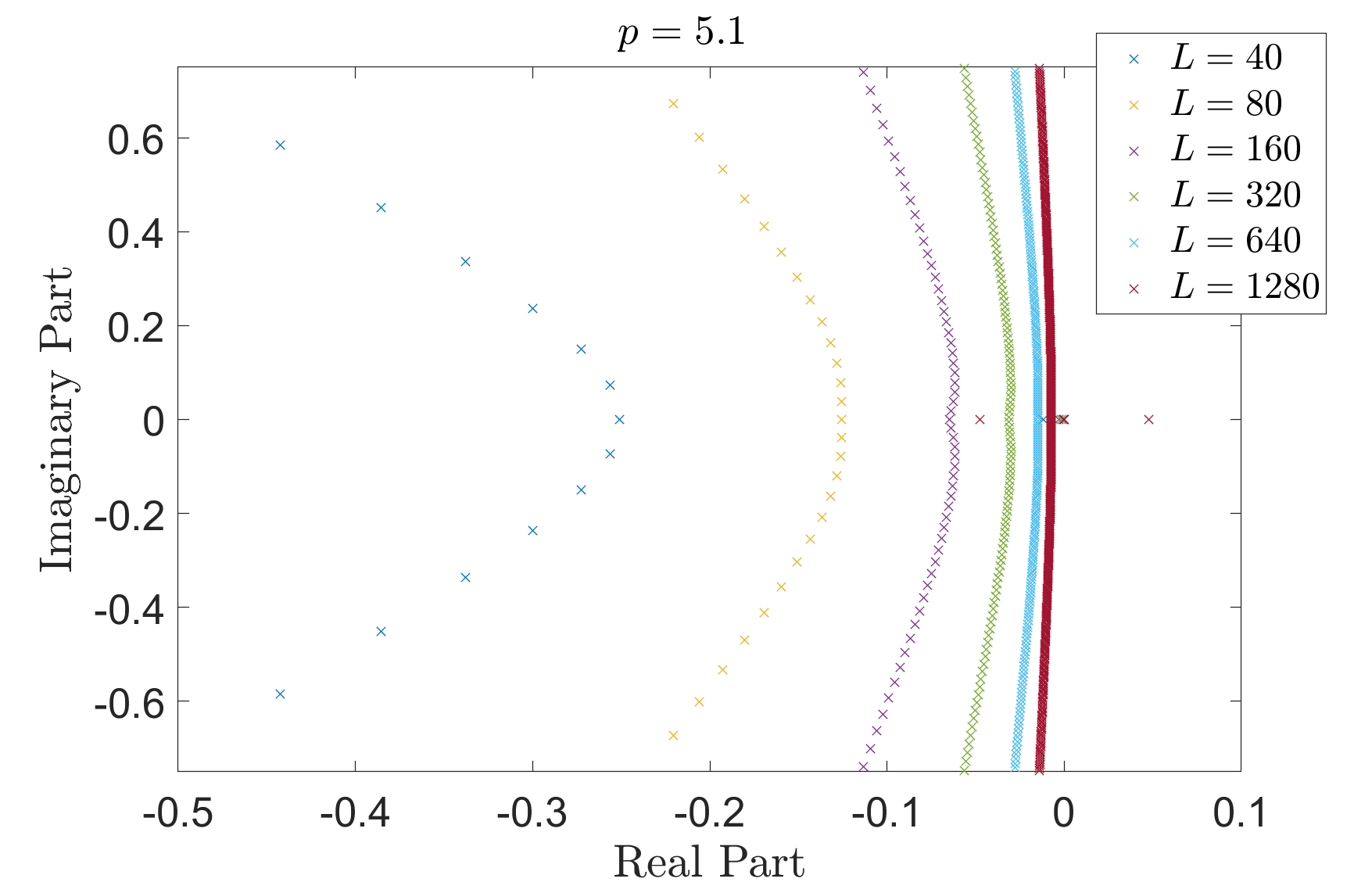}
\caption{Spectra of $\mathcal{L}$ [cf. Eq.~\eqref{eq:renKdV_eval_prob_op}]
associated with the TW branch of Fig.~\ref{fig:eigenvalue_soliton}. The
left and right panels of the figure showcase the full spectra of TWs
for $p=4.9$ (subcritical case) and $p=5.1$ (supercritical case), and
for various values of the domains' half-width $L$. We have performed
a convergence test of the spectrum by starting from $L=40$ and doubling
it each time up to $L=1280$ (see the legends therein in both panels).
As the infinite domain case is progressively approached,
the wedge-like spectrum in both panels starts aligning with an almost
vertical spectrum that is proximal to the imaginary axis.
\label{fig:soliton_subL_superL}}
\end{figure}

We finalize our discussion on the spectra of TW solutions to the gKdV by presenting their full spectra $\lambda=\lambda_{r}+\ii\lambda_{i}$ for two distinct values of $p$, considering various domain half-width, $L$, as depicted in Fig.~\ref{fig:soliton_subL_superL} (refer to the corresponding legends).
The left panel corresponds to the subcritical case of $p=4.9$, while the right panel pertains to the supercritical case of $p=5.1$, where the TW solution is unstable.
Some observations can be made based on the information presented in Fig.~\ref{fig:soliton_subL_superL}.
In both panels, the continuous spectrum is notably influenced by the size of the domain.
For relatively modest $L$ values, such as $L=40$ and $L=80$, the continuous spectrum appears to take on a wedge-like shape.
%
However, as the domain size increases, particularly for $L=640$ and $L=1280$, this shape transforms into a nearly vertical
(for the scale of the graph)
distribution approaching the imaginary axis and
thus suggesting the asymptotic form thereof.
%
%
In addition, the symmetry mode, accounting for
spatial translations, becomes increasingly distinguishable and well-resolved with larger $L$ values in our numerical computations, aligning with the expected behavior.
It is worth noting that the unstable mode that emerges for  $p>5$ (also observed in Fig.~\ref{fig:eigenvalue_soliton}) remains unaffected by changes in $L$.

\subsection{Existence of self-similar solutions: $G\neq 0$}\label{sec_3_2}

The observed instability of the TW as shown in Fig.~\ref{fig:eigenvalue_soliton}
for $p>5$ is consistent with the subsequent self-similar blow-up of the TW, a phenomenon already discussed in Sec.~\ref{sec_2}.
We now proceed with the computation of self-similar
solutions for the gKdV (Eq.~\eqref{eq:genKdV}).
As previously reported in Sec.~\ref{sec_2}, solutions that exhibit self-similarity in the renormalized / co-exploding frame translate to stationary solutions of Eq.~\eqref{eq:renKdV_v2}.
By directly simulating the initial-boundary value problem (IBVP) of Eqs.~\eqref{eq:renKdV_v2}, \eqref{eq:renKdV_BC_1}-\eqref{eq:renKdV_BC_2}
along with the pinning condition specified by Eq.~\eqref{pinning_cond},
$w(\xi,\tau)\mapsto w(\xi)$ as $\tau\rightarrow \infty$.
Here, $w(\xi)$ is a solution satisfying Eq.~\eqref{eq:renKdV_v3}.

\begin{figure}[pt]
\includegraphics[width = 0.99\textwidth]{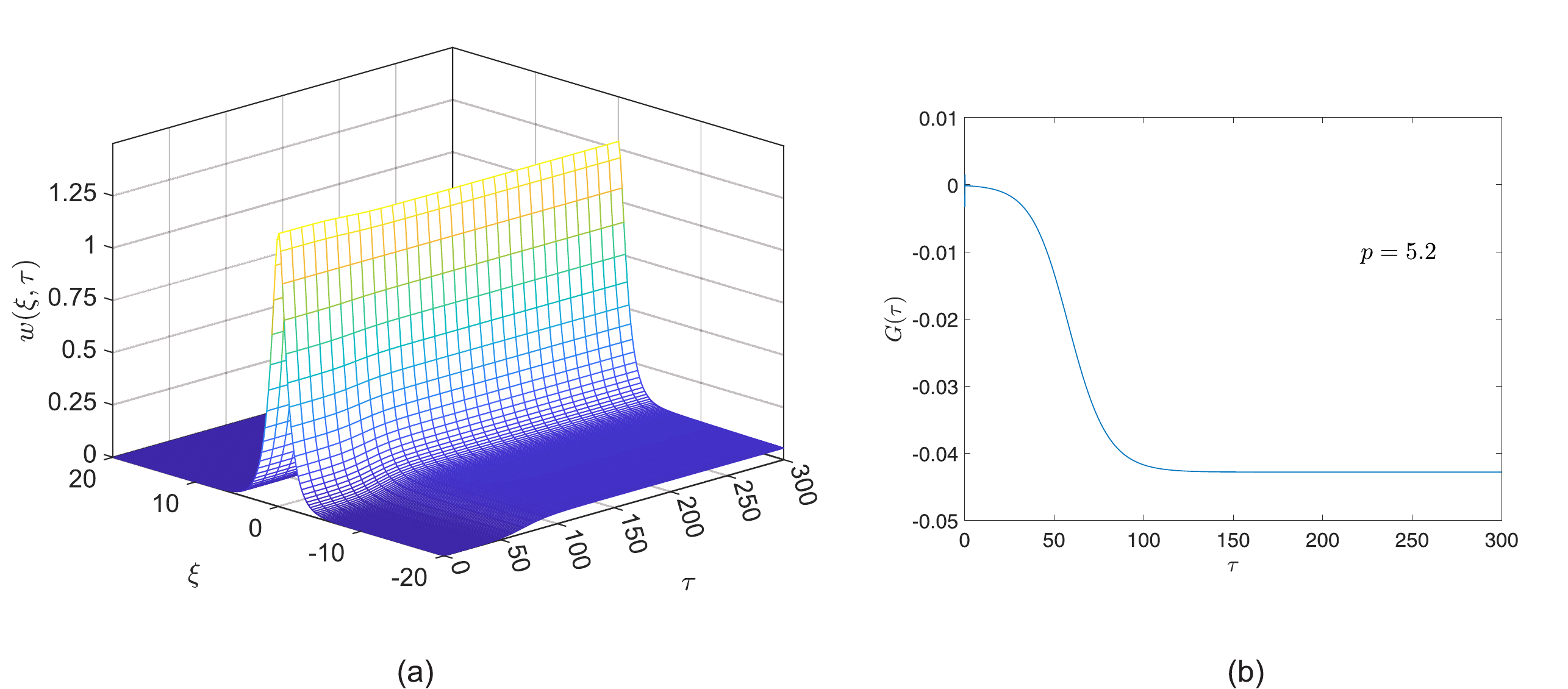}
\caption{Self-similar dynamics of Eq.~\eqref{eq:renKdV_v2} for
$p=5.2$. The TW solution of Eq.~\eqref{eq:soliton} was used
as an initial condition in the simulation. Panel (a) depicts
the evolution of the TW solution towards the self-similar profile
In panel (b), the temporal evolution of the blow-up rate $G(\tau)$
is monitored. Note that the self-similar dynamics converges to a
``steady-state'' solution of the co-exploding frame
when $\tau>100$.
\label{fig:self_similar_evolution}}
\end{figure}

This way, the self-similar profile can be determined by utilizing the TW solution as an initial condition, specifically for  a value of $p$ where the TW exhibits instability.
%
We integrate Eq.~\eqref{eq:renKdV_v2} in (the renormalized) time using a backward Euler method~\cite{Hairer}.
We have also tested our time integration algorithm by comparing its outcomes with those obtained from MATLAB's \textit{ode23t} function, showing a high degree of agreement.
Throughout the time integration process, we consistently impose the pinning condition defined in Eq.~\eqref{pinning_cond}
at each time step. This permits the computation of the rate of width change, $G$ (or blow-up rate) as a function of $\tau$, as illustrated in Eq.~\eqref{eq:A_func}.
The results of this simulation are depicted in Fig.~\ref{fig:self_similar_evolution}, specifically for $p=5.2$.
The left panel of the figure illustrates the spatio-temporal evolution
of $w(\xi,\tau)$ within the renormalized frame, while the right panel tracks the variation of the
blow-up rate, $G$, with $\tau$.
As evident from the panels, once a transient time window $\approx (0,100]$ elapses, the self-similar dynamics tends towards a stationary profile (as observed in the left panel).
Simultaneously, the blow-up rate, $G$, reaches a constant, negative value (as indicated in the right panel).
This stationary nature of the self-similar dynamics permits the accurate computation of the self-similar profile in accordance with Eq.~\eqref{eq:renKdV_v3}.
In particular, we extract the terminal solution at $\tau=300$ from the dynamics and employ it, in conjunction with the asymptotic value of $G$, as an initial guess for a Newton method, which rapidly converges. It is also important to highlight here
that the relaxational nature of the dynamics predisposes us
towards the expected dynamical stability of such a
self-similarly collapsing state.

The alternative, more computationally accurate approach adopted in our work for the numerical computation of self-similar solutions is by directly tackling the boundary value problem Eq.~\eqref{eq:renKdV_v3} with BCs of Eqs.~\eqref{eq:renKdV_BC_1}-\eqref{eq:renKdV_BC_2} and the pinning condition described in Eq.~\eqref{pinning_cond}.
%
Upon using either the above mentioned initial guess or
a slightly perturbed TW for $p>5$, along the most unstable eigenvector (as described in Eq.~\eqref{eq:stab_ansatz} - see also right panel of Fig.~\ref{fig:eigenvalue_soliton}),
we obtain the corresponding self-similar waveform for
the respective $p$.
Upon achieving convergence through the Newton's method
(either by employing this approach or the one mentioned earlier), we proceed with parametric continuation~\cite{kuznetsov_2023} across the nonlinearity exponent, $p$.
This step enables the full tracing of the branch of self-similar solutions.

\begin{figure}[pt]
\centering
\includegraphics[width = 0.495\textwidth]{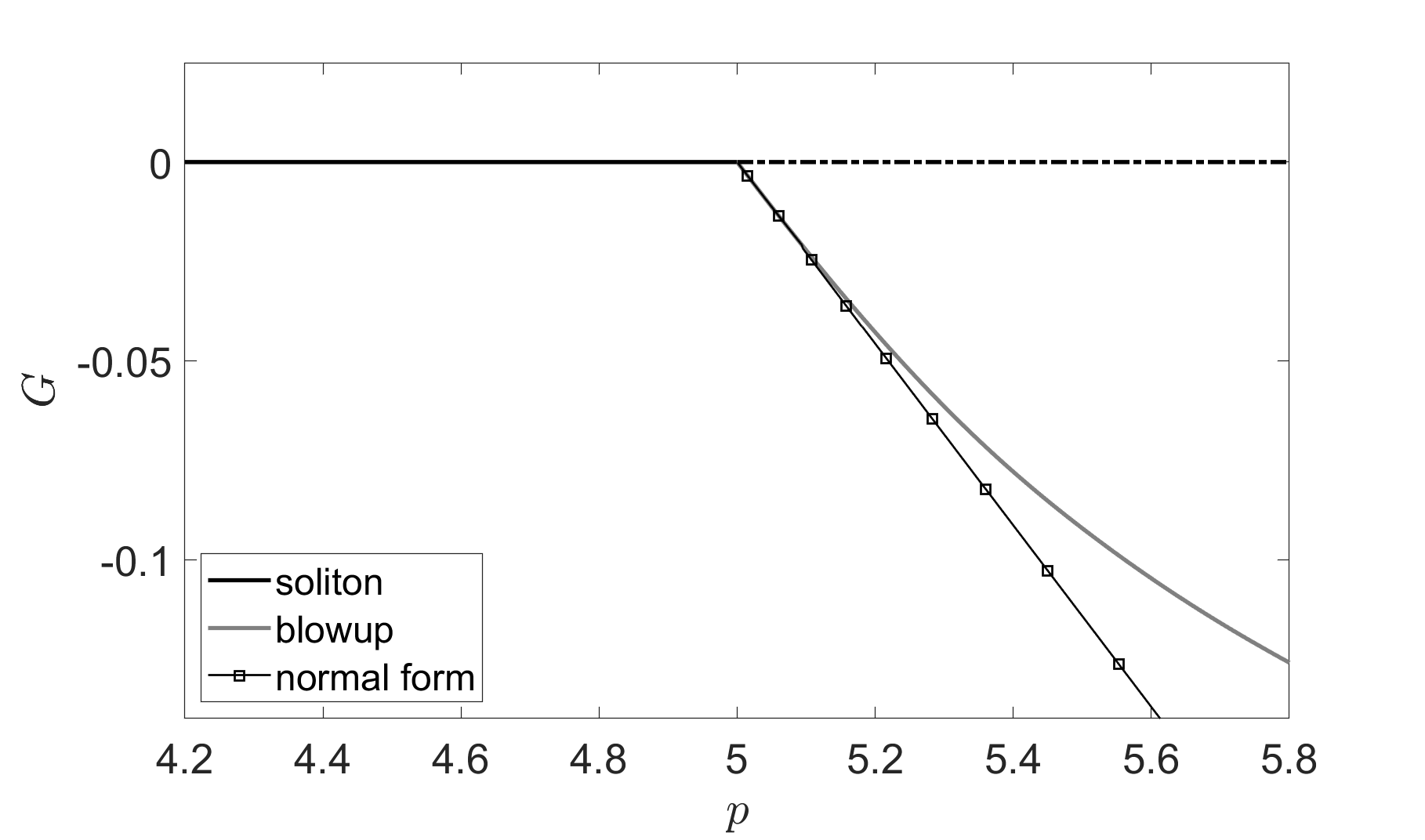}
\includegraphics[width = 0.495\textwidth]{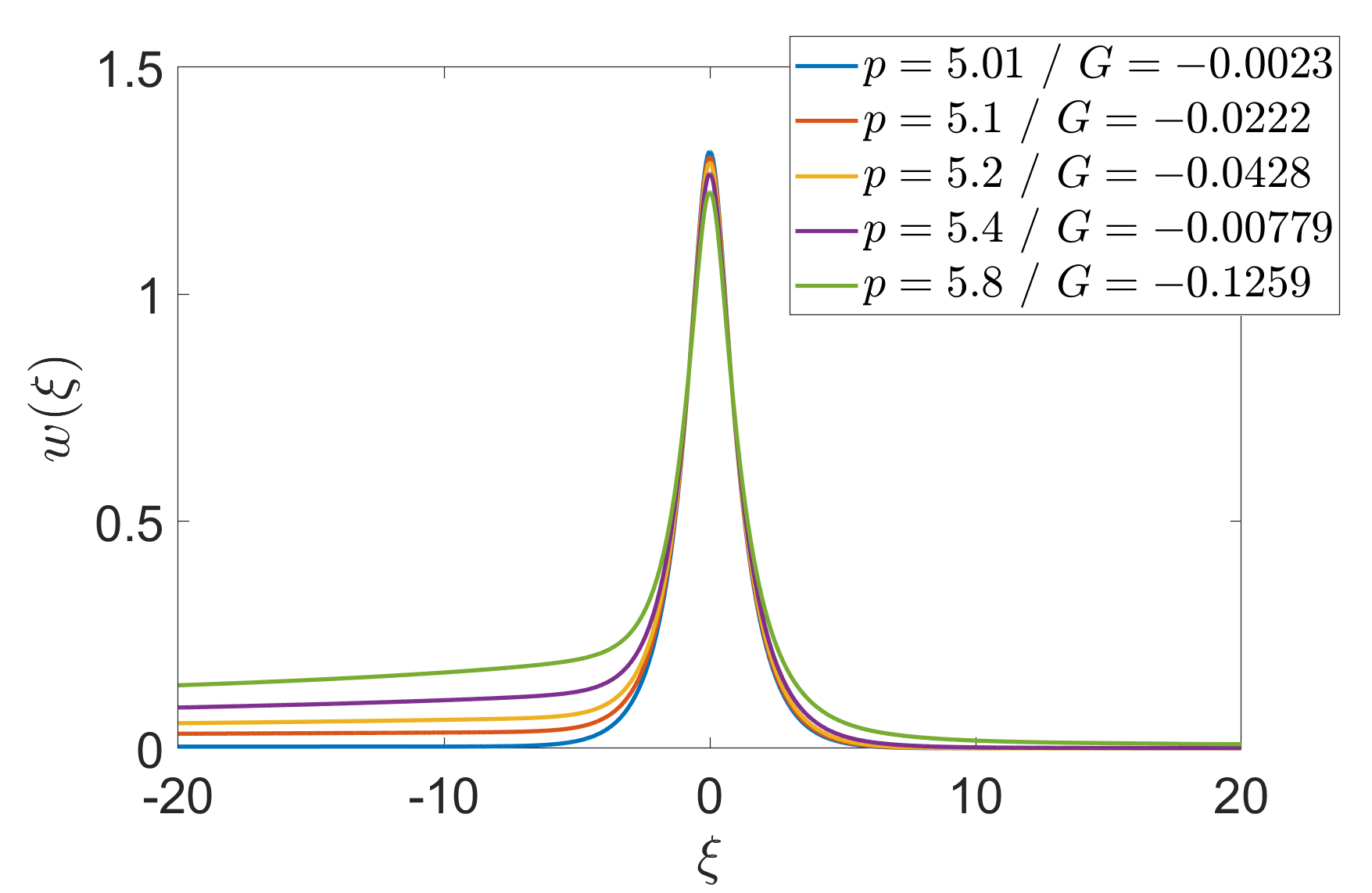}
\caption{Existence results of self-similar solutions to the gKdV obtained
numerically. In the left panel, the solid grey line corresponds to the
dependence of the blow-up rate $G$ on $p$ for the self-similar branch.
The black line with rectangles corresponds to the results obtained from the normal form (steady state solutions of Eq.~\ref{normalform}).
The solitonic branch with $G=0$ is overlaid in the panel, and shown with a solid black
line (for its stable portion), with the dashed part corresponding to the unstable segment of the solitonic branch.
The right panel presents a few self-similar profiles for different values
of $p$ (and $G$). For their values, see the legend therein.
\label{fig:blow-up_bifdiagram}}
\end{figure}

Displayed in Fig.~\ref{fig:blow-up_bifdiagram} are our numerical findings regarding the existence of self-similar solutions to the gKdV equation (as defined in Eq.~\eqref{eq:renKdV_v3}).
In the left panel of the figure, the dependence of $G$ on $p$
for the self-similar branch ($G<0$) is illustrated by the grey line, while the solid black line represents the soliton (TW) branch.
The dashed black line segment signifies the unstable nature of
the solitary wave solutions for $p>5$.
On the right panel, the numerically obtained self-similar profiles for various $p$ values are presented, along with the corresponding blow-up rates, $G$, indicated in the legend.
Notably, the profiles presented in this panel exhibit identical characteristics to those reported in~\cite{weinmueller_2020}.
Specifically, the solutions display a power law decay on the left side and a rapid (exponential) decay on the right side, with respect to the peak located at the origin.
Through the application of the transformation $G\mapsto-G$ and $\xi\mapsto-\xi$,
we observe that the regions capturing the asymptotic behaviors of the solutions are indeed exchanged (results not displayed).
The resulting profiles align with those presented in~\cite{koch_2015} under this transformation.
%

\subsection{Spectral analysis of self-similar solutions: $G<0$}\label{sec_3_3}

We now shift our focus to the spectral analysis of the self-similar solutions within the context of the gKdV equation.
In this pursuit, we recall the eigenvalue problem outlined in Eq.~\eqref{eq:renKdV_eval_prob}, along with its corresponding operator, $\mathcal{L}$, as presented in Eq.~\eqref{eq:renKdV_eval_prob_op}.
During each step of the continuation method, which we have  employed to trace
the self-similar branch ($G<0$), we compute the spectrum of the operator,
$\mathcal{L}$.
Our ensuing analysis will be divided into two components:
one centered around the point spectrum, and the other concerning the continuous spectrum.
Regarding the point spectrum, our findings reveal the existence of four real
eigenvalues that can be systematically traced.
Among these four, two eigenvalues are positive, denoted as $\lambda_{1}$ and $\lambda_{2}$, with
$\lambda_{1}>\lambda_{2}$.
The remaining two eigenvalues are negative, designated as $\lambda_{-1}$ and $\lambda_{-2}$, with $\lambda_{-1}>\lambda_{-2}$.

\begin{figure}[pt]
\centering
\includegraphics[width = 0.75\textwidth]{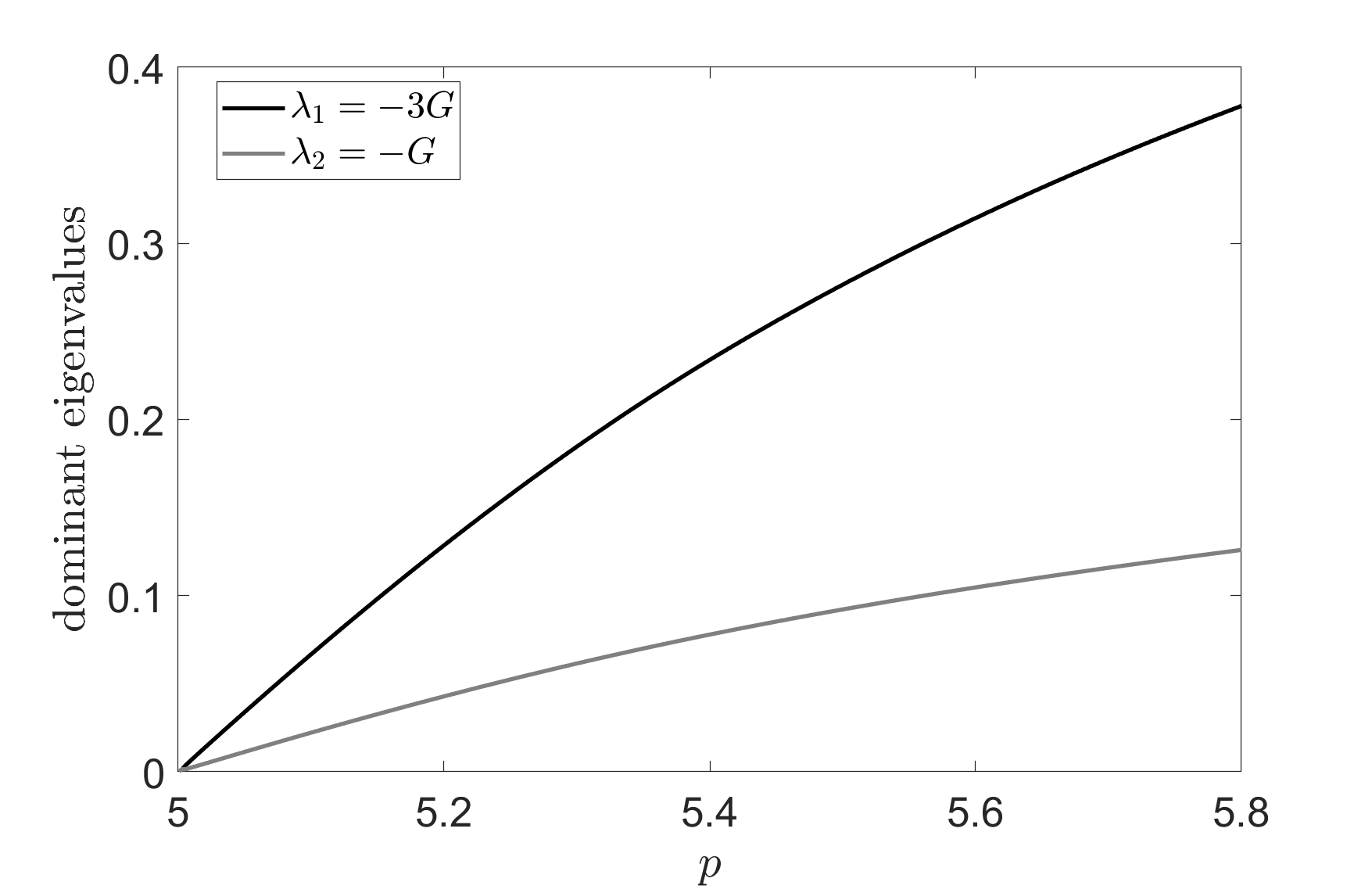} 
\caption{The dependence of the 2 unstable eigenvalues as a function of $p$.
Note that those are precisely located at $\lambda_1=-3G$ and $\lambda_2=-G$.
\label{fig:dominanteigs}}
\end{figure}

Let us begin our analysis with the two positive eigenvalues that pertain to {\it apparent} instabilities in the co-exploding frame.
We find that these eigenvalues are
precisely situated at $\lambda_{1}=-3G$ and $\lambda_{2}=-G$, a fact readily observed in Fig.~\ref{fig:dominanteigs}. This is
true up to exponentially small corrections in the size of
the domain which are imperceptible over the scales shown
and for the choices of $L$ made herein.
%
Accompanying these eigenvalues, we also present the explicit forms of their associated eigenfunctions (for the infinite domain
problem).
Specifically, the eigenvector associated with $\lambda_{1}=-3G$ is:
\begin{align}
\label{eq:evec_1}
v(\xi)=\frac{2}{p-1}w+\xi w_{\xi}+\frac{1}{G}w_{\xi},
\end{align}
where the subscript indicates differentiation with respect to $\xi$.
Similarly, when considering $\lambda_{2}=-G$ the (exact, in
the infinite domain limit) eigenvector takes the form:
\begin{align}
\label{eq:evec_2}
v(\xi)=w_{\xi}.
\end{align}
This latter eigenvector is associated with the translational
invariance of the problem, while the former is connected
with the scaling symmetry and the associated invariance.
Fig.~\ref{fig:domeigvector} presents a comparison between the exact eigenvectors derived from Eqs.~\eqref{eq:evec_1}-\eqref{eq:evec_2} (depicted as solid black lines) and the numerically computed counterparts (depicted as gray dots).
For the sake of illustration, we choose a specific value for the nonlinearity exponent, namely $p=5.2$.
It is evident from the panels of the figure that the two sets of eigenvectors exhibit a remarkable agreement with their
corresponding theoretical prediction for the choices of
$L$ depicted herein.

Let us now delve into the discussion concerning the two aforementioned negative real eigenvalues, namely $\lambda_{-1}$ and $\lambda_{-2}$, which we systematically trace during our computations.
The eigenvalue, $\lambda_{-1}$, aligns fairly closely with
$G/2$, while $\lambda_{-2}$ displays oscillations centered around $2G$.
The left panel of Fig.~\ref{fig:dominanteigs_negative} illustrates these behaviors, depicting the dependence of these negative real eigenvalues on the nonlinear exponent, $p$ with  solid blue (accompanied by open circles) and red (with open squares) lines,
respectively.
In the same panel, dashed-dotted red and blue lines correspond to the values of $G/2$ and $2G$ as functions of $p$, included for comparison.
Examining the left panel, we observe that while the eigenvalue, $\lambda_{-1}$, remains consistently proximal to the
prediction of $G/2$ (see also details below) starting from its emergence at $p=5$, it does exhibit small oscillations that gradually increase in amplitude as $p$ grows.
Similarly, the eigenvalue, $\lambda_{-2}$, displays more vigorous
oscillatory behavior around $2G$ (indicated by the solid red line with open squares).
Furthermore, the panel showcases the behavior of the third and fourth largest negative eigenvalues, denoted as $\lambda_{-3}$ and $\lambda_{-4}$, depicted with solid
orange (with open triangles) and purple (with stars) lines, respectively.
Similar to $\lambda_{-2}$, these eigenvalues also experience oscillations around $2G$, which amplify as $p$ increases.
It is worth noting that these oscillations are a consequence of the finite domain size employed in our analysis, and we will delve into a systematic analysis of these oscillations later.
Illustrating the effect of the domain size, the right panel of Fig.~\ref{fig:dominanteigs_negative} compares the dependence of
$\lambda_{-4}$ on $p$ for two different domain half-widths, specifically $L=40$ and $L=160$.
The solid blue line (with open circles) and orange line (with open circles) represent $\lambda_{-4}$ for $L=40$ and $L=160$, respectively.
This comparison highlights that as the domain half-width increases
from $40$ to $160$, the real part of $\lambda_{-4}$ decreases (in terms of its absolute value) from its initial value at $p=5$.
Additionally, the oscillations in the eigenvalue are discernible at lower values of $p$ for the larger domain size ($L=160$), though $\lambda_{-4}$ continues to oscillate around $2G$ in both cases.

\begin{figure}[pt]
\centering
\includegraphics[width = 0.49\textwidth]{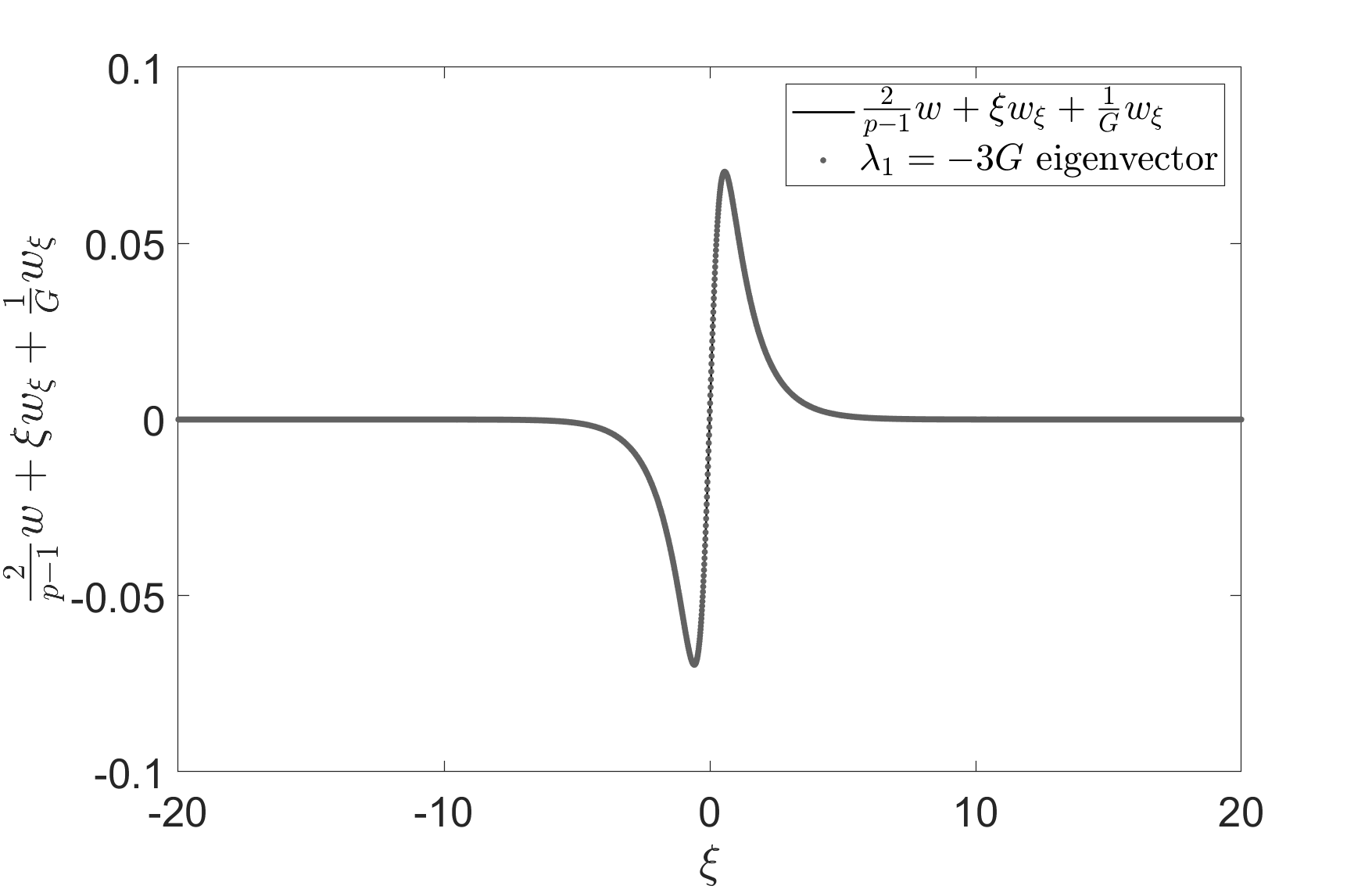}
\includegraphics[width = 0.49\textwidth]{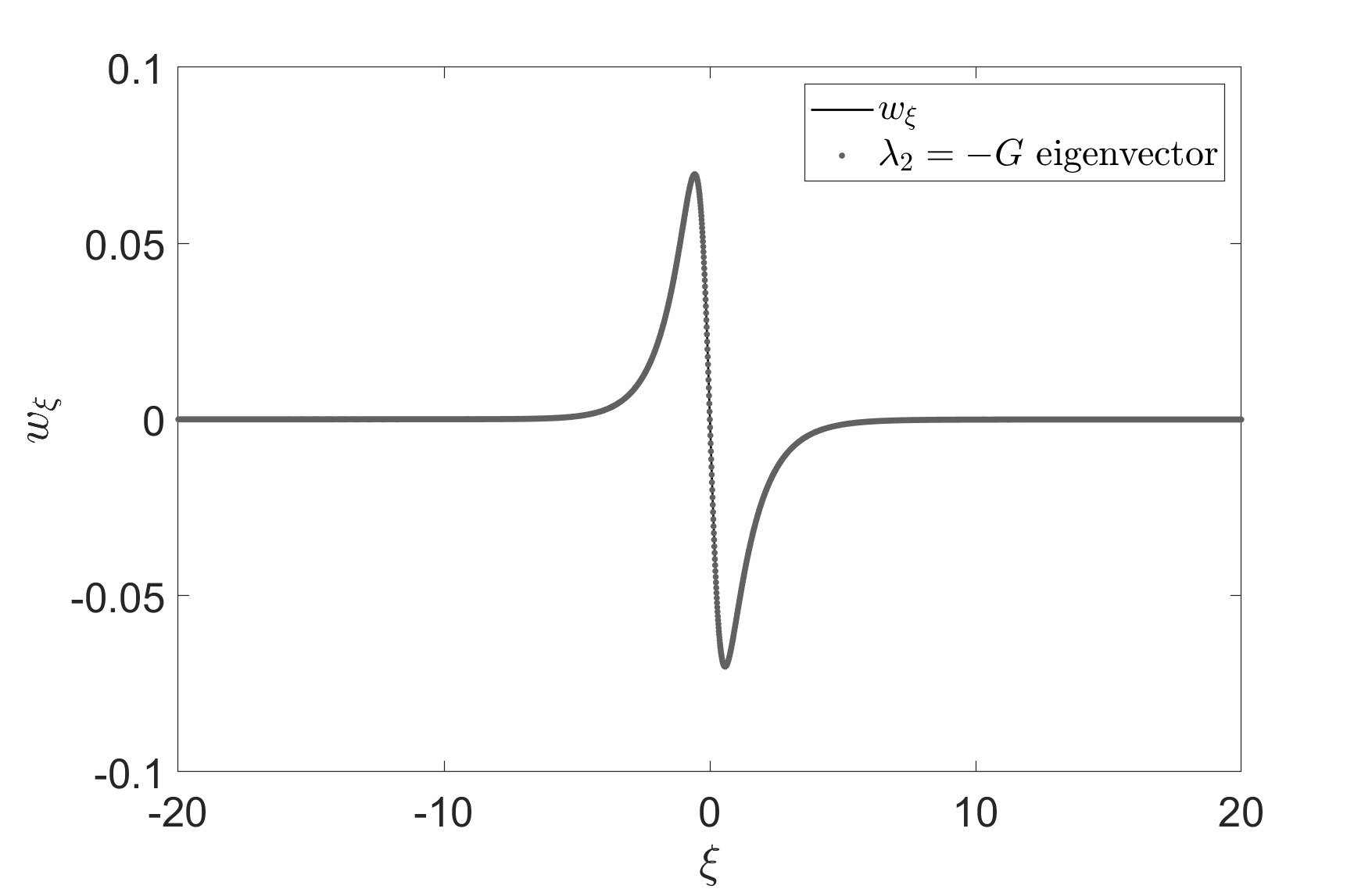}
\caption{
Comparison of (the practically identical) exact and numerically obtained eigenvectors
shown with a black solid line and grey dots, respectively,
for $p=5.2$. The left panel presents the eigenvector
associated with $\lambda_{1}=-3G$ and the scaling invariance [cf. Eq.~\eqref{eq:evec_1}]
whereas the right panel depicts the one associated with $\lambda_{2}=-G$, associated with translational invariance
(cf. Eq.~\eqref{eq:evec_2}).
\label{fig:domeigvector}}
\end{figure}
\begin{figure}[pt]
\centering
\includegraphics[width = 0.49\textwidth]{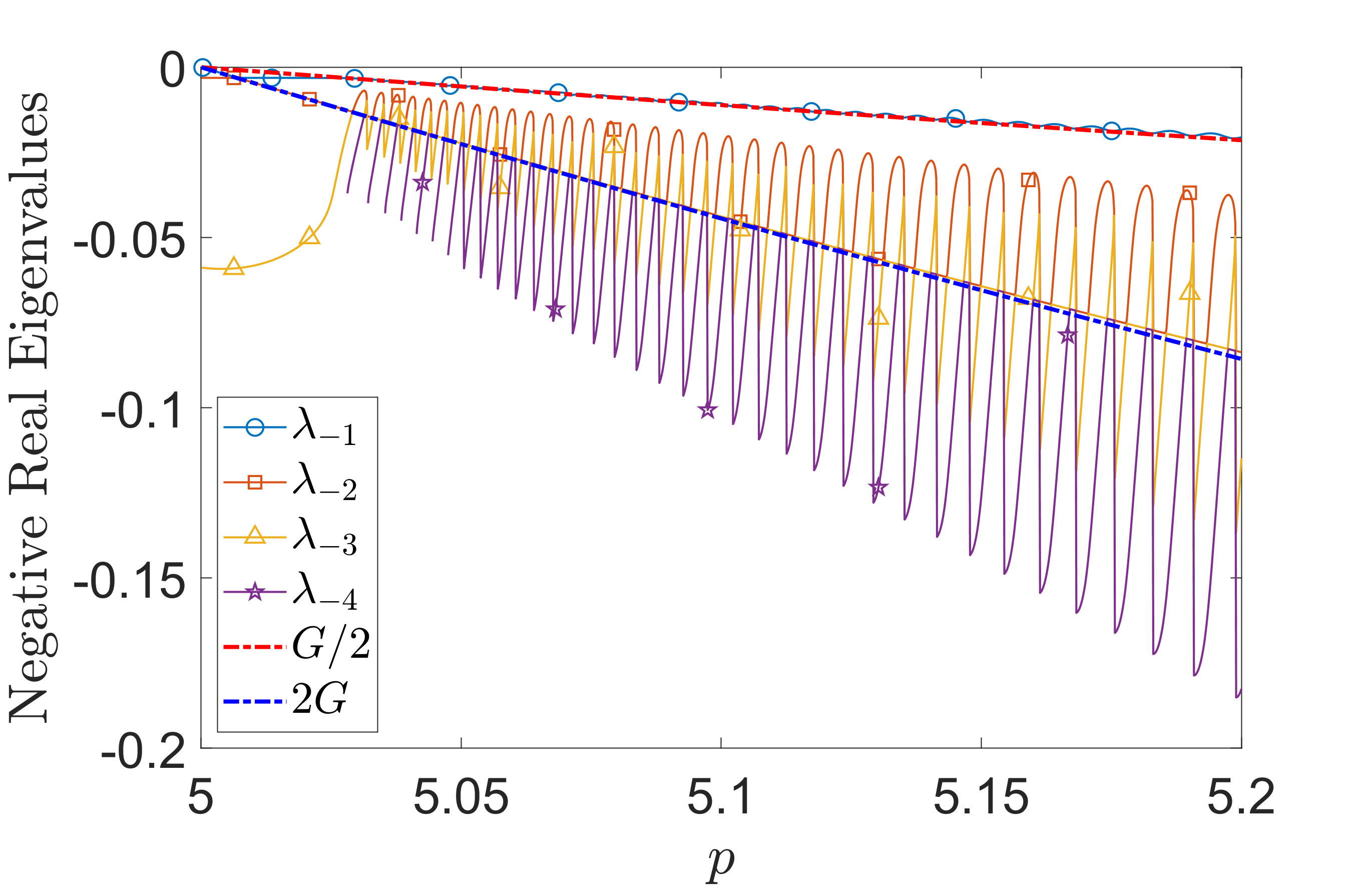}
\includegraphics[width = 0.49\textwidth]{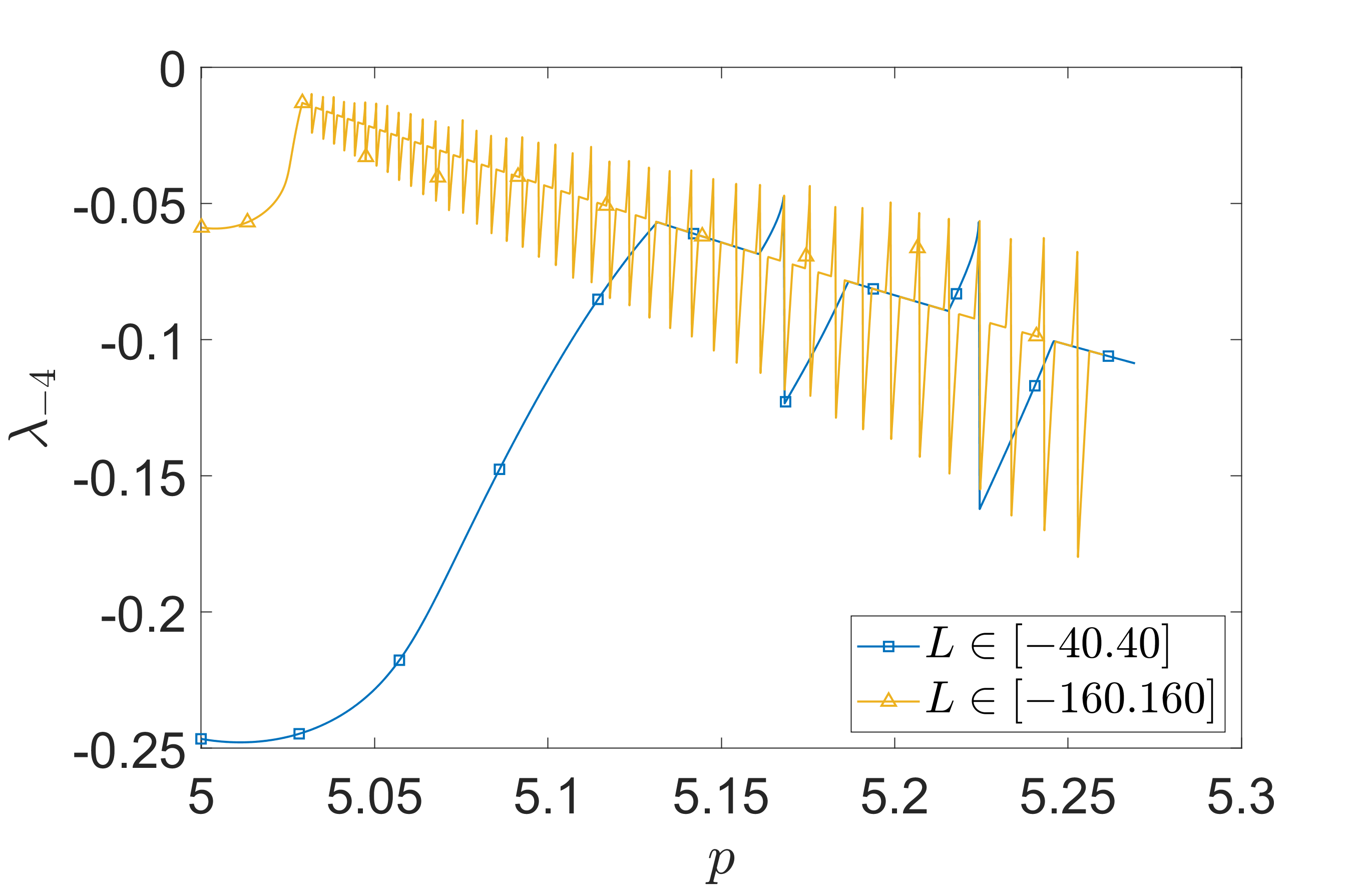}
\caption{Continuation of Fig.~\ref{fig:dominanteigs} but now for the
largest negative real eigenvalues. The left panel presents the variation
of the 4 largest negative real eigenvalues,
$\lambda_{-1}>\lambda_{-2}>\lambda_{-3}>\lambda_{-4}$ over $p$, and
for $L \in [-160,160]$ (see the legend therein). In particular, $\lambda_{-1}$
oscillates around $G/2$ (red dashed-dotted line) whereas $\lambda_{-2,-3,-4}$
around $2G$ (depicted with blue dashed-dotted line). The right panel highlights
the finite-size domain effect by showcasing the variation of $\lambda_{-4}$
over $p$, and for $L \in [-40,40]$ (solid blue line with open circles) and for
$L\in[-160,160]$ (solid orange line with open triangles).
\label{fig:dominanteigs_negative}}
\end{figure}

\begin{figure}[pt]
\centering
\includegraphics[width = 0.49\textwidth]{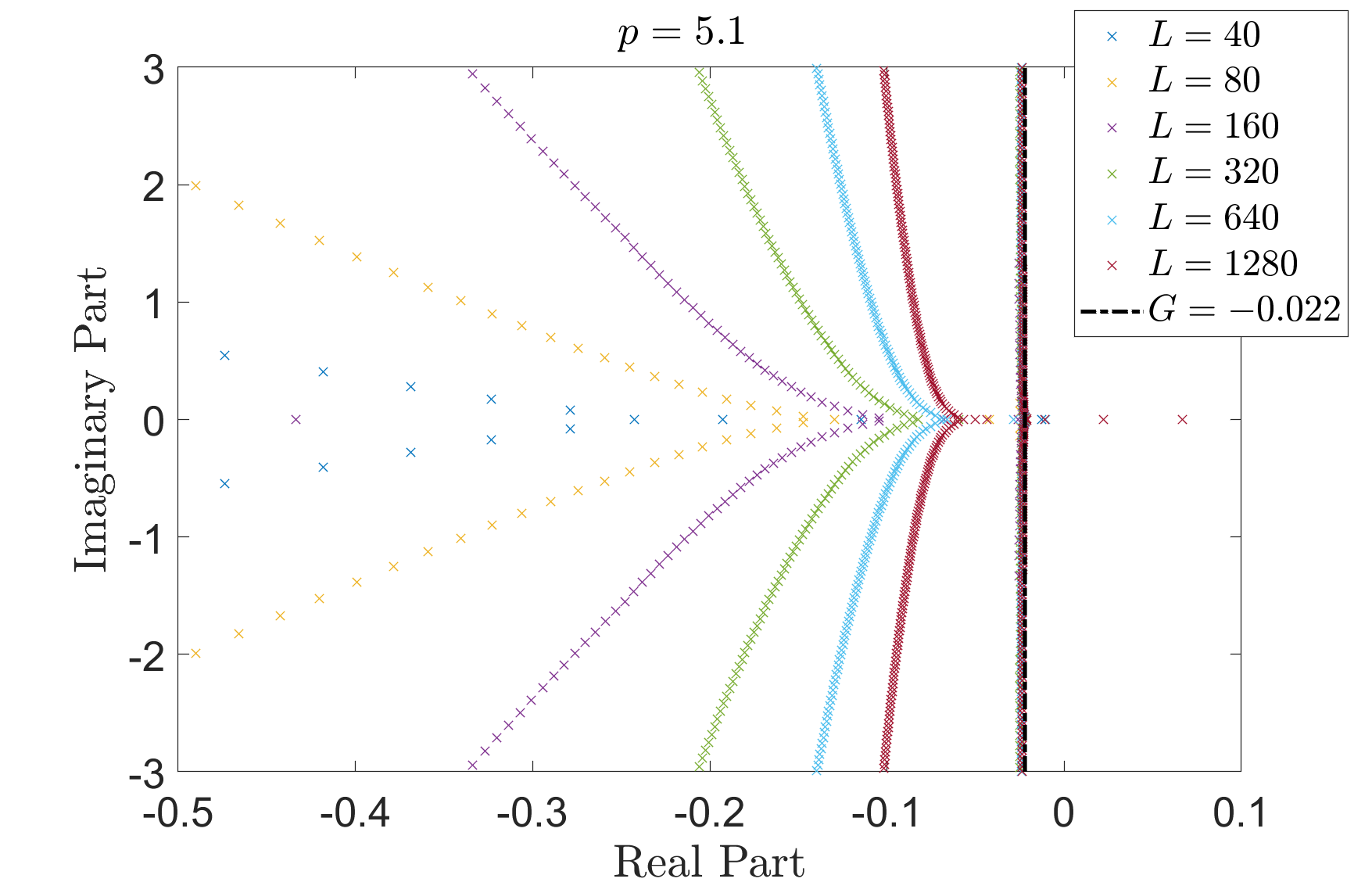}
\includegraphics[width = 0.49\textwidth]{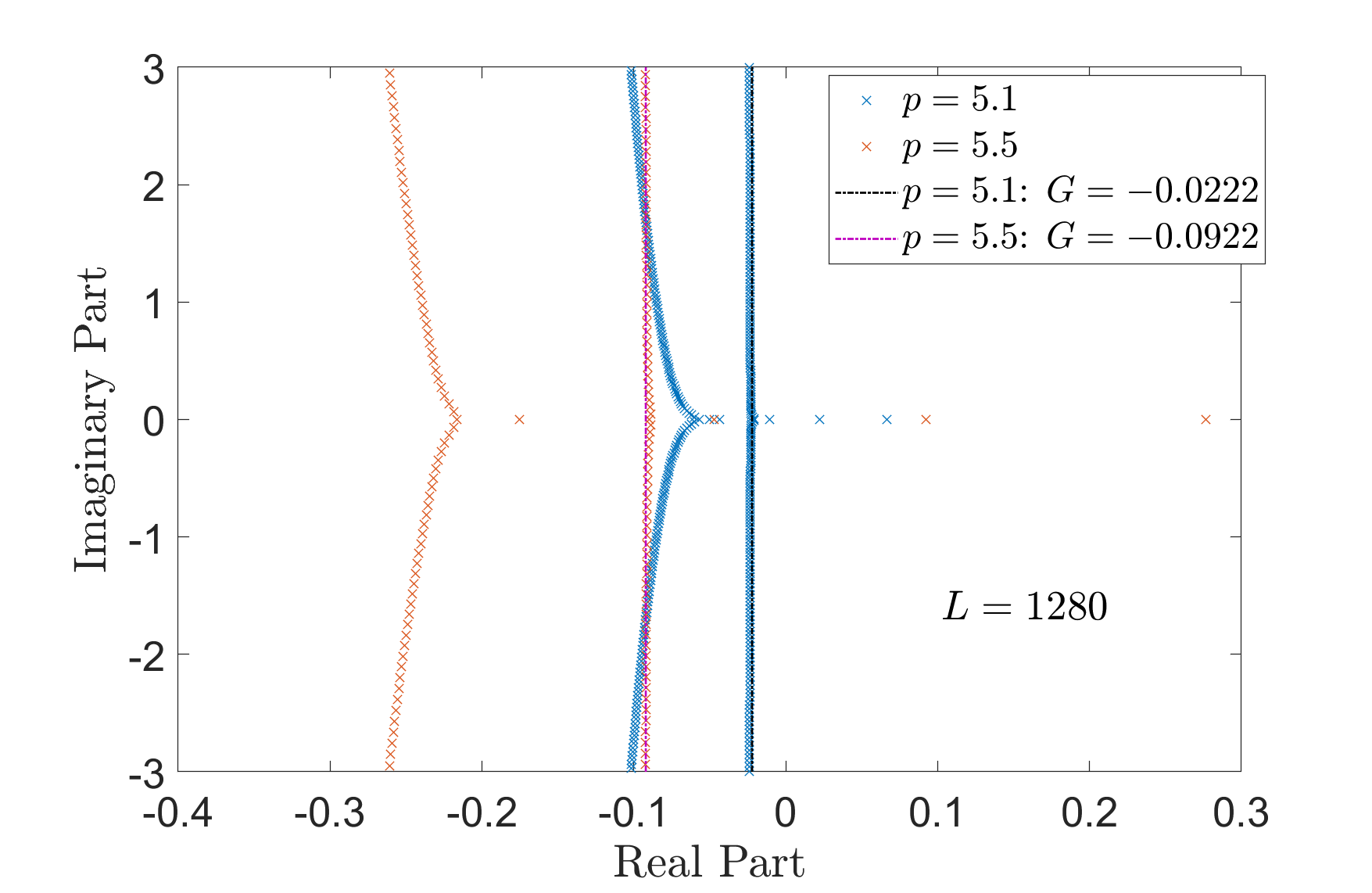}
\caption{Same as Fig.~\ref{fig:soliton_subL_superL} but for the
self-similar branch. In the left panel, we present spectra of
$\mathcal{L}$ [cf. Eq.~\eqref{eq:renKdV_eval_prob_op}] with $p=5.1$
as functions of the domain size $L$ (see the legend therein).
We employ a similar convergence analysis over $L$ as we performed
in Fig.~\ref{fig:soliton_subL_superL}. Interestingly, the spectrum
involves two parts: a vertical part that lands exactly on $G\approx -0.022$
(see also the dashed-dotted line, representative of this value), and
a wedge-like one that is morphed into a vertical spectrum as $L$ increases,
i.e., as we approach the infinite
domain case; yet this portion of
the spectrum will be argued to not
be well-resolved in what follows.
The right panel of the figure,
considers $L=1280$, and compares the spectrum of two self-similar solutions
with $p=5.1$ (blue crosses) and $p=5.5$ (red crosses). The dashed-dotted
black and purple lines are introduced to highlight the location of the
vertical parts of the spectrum at $\mathrm{Re}(\lambda)\approx G$
(see the legend).
\label{fig:spectraLP}}
\end{figure}

The finite-size effects,
whose influence we have already seen in part,
leave their imprint on the continuous spectrum of self-similar solutions
as well.
The left panel of Fig.~\ref{fig:spectraLP} depicts the spectrum of the operator $\mathcal{L}$ [cf.~Eq.~\eqref{eq:renKdV_eval_prob_op}] for a specific value of $p=5.1$, as it varies with the domain's half-width, $L$ (again, by doubling $L$ each time as we performed in Fig.~\ref{fig:soliton_subL_superL}).
However, the continuous spectrum in this context presents distinctive attributes compared to the TW case.
The continuous spectrum can be split into two segments: an almost-vertical portion situated close to $G\approx -0.022$, and a wedge-like portion, reminiscent of the one observed in the TW case.
Upon increasing $L$, we note that the nearly-vertical segment of the continuous spectrum remains relatively consistent, gradually converging toward a real part of the relevant spectrum equal to $G$.
It is relevant to note here that this feature is
reminiscent of the continuous spectrum in the
NLS case~\cite{chapman_2022}.
Concurrently, the wedge angle associated with the other portion of the continuous spectrum widens, leading to a transformation of the wedge-like section into an almost-vertical line parallel to the imaginary axis.
Meanwhile, the distance between this almost-vertical section of the spectrum and the imaginary axis expands as $p$ increases.
This is illustrated in the right panel of Fig.~\ref{fig:spectraLP}, where a comparison of the spectra between $p=5.1$ and $p=5.5$ is presented using blue and red crosses, respectively.
Nevertheless, as will be discussed
below, this portion of the spectrum
is {\it not} well-resolved. More
specifically, we will argue that this
segment is
{extremely ill-conditioned} and,
in reality, it is supposed to lie on the
negative real axis.
It is noted that, similarly to the $p=5.1$ case, the vertical section of the spectrum at $p=5.5$ aligns precisely with $G\approx -0.0922$.

\subsubsection{Spectral analysis of self-similar solutions using WKB}\label{sec_3_3_1}

We now proceed with the analytical approximation of both the continuous and point spectrum of the operator, $\mathcal{L}$, by using the Wentzel-Kramers-Brillouin (WKB) method~\cite{bender_orszag_book}.
We suppose that we are close to the bifurcation point, so that $p$ is close to 5 and $G$ is close to zero. We will see in Section \ref{sec_3_4} that the steady self-similar solution comprises an inner region in which $\xi = O(1)$, and an outer region in which $\xi = O(1/\vareps)$, where $\vareps = -G>0$ measures the distance from the bifurcation point, so that $p-5$ is proportional to $\vareps$. This inner-outer structure is inherited by the eigenvalue problem. However, we will find that the spectral properties are mainly determined by the outer region.
Thus we adopt a rescaling into the outer region by introducing the transformations:
\begin{align}
\label{eq:resc_outer}
y = \vareps\,\xi, \quad G = -\vareps, \quad \vareps>0.
\end{align}
After substituting Eqs.~\eqref{eq:resc_outer} into
Eqs.~\eqref{eq:renKdV_eval_prob}-\eqref{eq:renKdV_eval_prob_op}, we
obtain
\begin{align}
\label{eq:leading_0}
\frac{\lambda}{\vareps}v=-\vareps^{2}\frac{\d^{3}v}{\d y^{3}}%
-p\frac{\d}{\d y}{\left(vw^{p-1}\right)}-\left(\frac{2v}{p-1}+y\frac{\d v}{\d y}\right)
+\frac{\d v}{\d y}.
\end{align}
Since $w^{p-1}$  is exponentially small in the
outer region it may be neglected. Expanding $p$ as $p=p_{0}+\vareps\,p_{1}+\cdots$ (with $p_{0}=5$),
we obtain:
\begin{align}
\frac{\la}{\vareps} v + \vareps^2 \frac{\d^3 v}{\d y^3}-\frac{\d v}{\d y}=%
-\left(\frac{v}{2}  + y \frac{\d v}{\d y}\right) +\frac{\vareps p_1 v }{8} +\cdots.
\label{eq:leading}
\end{align}
Subsequently, we introduce the WKB ansatz:
\begin{align}
v \sim A \ee^{\phi/\vareps}, \quad \vareps \downarrow 0,
\end{align}
where $\phi(y)$ represents the phase and $A(y)$ corresponds to the slowly varying amplitude.
Equating coefficients of powers of $\vareps$ at orders $\mathcal{O}(\vareps^{-1})$ and $\mathcal{O}(1)$ gives, respectively,
\begin{subequations}
\begin{align}
\la + (\phi')^3 +(y-1) \phi' = & \,\,0,
\label{eq1_WKB}
\\
3 A' (\phi')^2 + 3 A \phi' \phi'' - A' = &\,\, -\frac{A}{2} -y A',
\label{eq2_WKB}
\end{align}
\end{subequations}
the latter of which can be further rearranged as
\begin{align}
\frac{ A' }{A} + \frac{1}{2}\frac{6  \phi'\phi''+1}{3 (\phi')^2-1+y} =0,
\end{align}
and subsequently integrated with respect to $y$, yielding
\begin{align}
\label{eq:sol_A}
A = \frac{1}{(3 (\phi')^2-1+y)^{1/2}},
\end{align}
where we have set the constant of integration to unity.
Since Eq.~\eqref{eq1_WKB}
is a cubic there are three distinct WKB solutions, which  we denote
by  $v_i \sim A_i  \ee^{\phi_i/\eps}$,
with $i = 1,2, 3$.
We choose these so that for real $y$, the phase functions have asymptotic behaviour
\begin{align}
\phi_1' \sim  (-y)^{1/2}, \qquad  \phi_2' \sim - (-y)^{1/2}, \qquad
\phi_3' \sim - \frac{\la}{y} \qquad \mbox{ as } y \ra -\infty.
\label{eq:WKB_approxs}
\end{align}
We find that there are  turning points  at
\begin{align}
\label{eq:turning_points}
y = 1 - 3 \left(\frac{\la}{2}\right)^{2/3}.
\end{align}
Because of the third root there are three turning points, which we label as $y_A$, $y_B$ and $y_C$.

Naively  we might now try to approximate an eigenfunction as a linear
combination of $v_1$, $v_2$ and $v_3$. However, in doing so it is
crucial to understand both (i) how the WKB solutions are permuted when
circling a turning point because of the Riemann surface associated
with the function $\phi$ (turning points are branch points of $\phi$),
and, (ii) how the coefficients in such  a linear combination will
change discontinuously across the Stokes lines associated with each of
the turning points \cite{Meyer,ParisWood,OldeDaalhuis}.
The following analysis of the eigenvalue problem uses a similar
approach to that in
\cite{transition}.

\begin{figure}[pt]
\begin{center}
\begin{overpic}[width=0.6\textwidth]{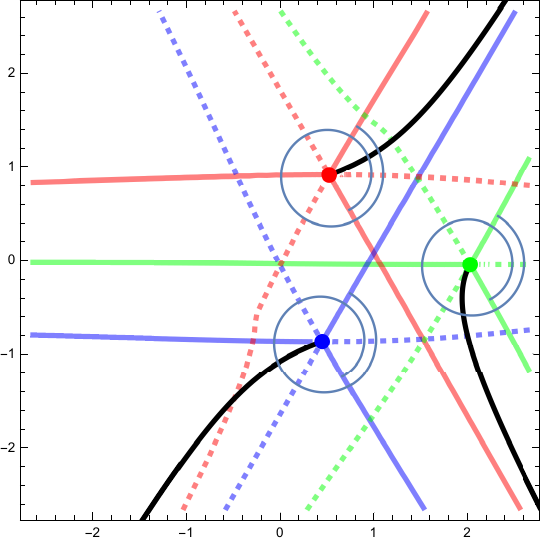}
\put(57,64){$y_A$}
\put(58,32){$y_B$}
\put(86,52){$y_C$}
\put(50,60){\small $\phi_1$}
\put(63,64){\small $\phi_3$}
\put(70,63){\small $\phi_1$}

\put(47,40){\small $\phi_3$}
\put(63,38){\small $\phi_3$}
\put(69,38){\small $\phi_2$}

\put(74,48){\small $\phi_1$}
\put(90,49){\small $\phi_1$}
\put(95,44){\small $\phi_2$}

\put(10,33){\small Stokes line}
\put(27,93){\rotatebox{-65}{\small Anti-Stokes line}}
\put(26,9){\rotatebox{55}{\small branch cut}}

\put(50,-5){$\re(y)$}
\put(-5,50){\rotatebox{90}{$\im(y)$}}
\end{overpic}
\end{center}
\caption{Stokes lines (solid) and anti-Stokes lines (dashed)
when $\la \approx -0.02537 - 0.401\ii$. Solid black lines are
branch-cuts of the phase function $\phi$. Three turning points $y_A$,
$y_B$ and $y_C$ are indicated. Each is a square root branch point of
$\phi$. The local Riemann surface structure in the vicinity of each
branch point is indicated by the spirals, which shows which branches
of $\phi$ are turned on/off across the Stokes lines.}
\label{fig:StokesA}
\end{figure}

A typical Stokes line picture is shown in Fig.~\ref{fig:StokesA}.
When the turning point $y_A$ is encircled, the phase $\phi_1$ becomes $\phi_3$ and vice versa.
Similarly, as the turning point $y_B$ is encircled, $\phi_2$ becomes  $\phi_3$,
and vice versa, while when the turning point $y_C$ is encircled, $\phi_1$ switches places with
$\phi_2$.
For definiteness we  introduce branch cuts
across which these interchanges in labels take place (see Fig.~\ref{fig:StokesA}).
Stokes lines in Fig.~\ref{fig:StokesA} are indicated by solid lines. Across each of the lines a dominant WKB solution will switch on a subdominant WKB solution (i.e. there will be a change in the coefficient of the subdominant solution proportional to the coefficient of the dominant solution). The relevant dominant/subdominant WKB solutions depend on the turning point at which the Stokes lines originate. Also shown in Fig.~\ref{fig:StokesA} are anti-Stokes lines (dashed) across which the dominance of two WKB solutions switches.
Finally, the spirals around each turning point illustrate one of the possible local branch point/Stokes switching structures. Let us explain these by considering the diagram in the vicinity of the turning point $y_C$, for example, with a solution in which only the WKB solution $v_1$ is present to the left of the turning point. There is a possible Stokes line close to the negative real axis but it is not active because $\phi_1$ is subdominant to $\phi_2$ there. As the turning point is encircled clockwise we first cross the anti-Stokes line at which $\phi_1$ becomes the dominant exponential, followed by another Stokes line. Since $\phi_1$ is now dominant, a multiple of $\phi_2$ is turned on. Proceeding clockwise we cross another anti-Stokes line close to the positive real axis across which $\phi_1$ and $\phi_2$ again exchange dominance. We then meet another Stokes line, at which $\phi_2$ is the dominant exponential. At this Stokes line there is a change in the coefficient of $v_1$---in fact the change is such that the coefficient is zero after the Stokes line is crossed, so that only $\phi_2$ remains. We then cross the branch cut, so that $\phi_2$ becomes $\phi_1$ and we return to where we started.

The approximate eigenvalue condition stipulates that the anti-Stokes line emanating from one of the turning points intersects the right-hand
boundary (see Fig.~\ref{fig:StokesA}, for which the anti-Stokes line from $y_C$ passes through the right-hand boundary).
This gives three branches of eigenvalues (one for each turning point), which are roughly (A)  $\re(\la)\approx G$, $\im(\la)>0$; (B) $\re(\la) < G$, $\im(\la)=0$; and (C)
$\re(\la)\approx G$, $\im(\la)<0$.

To see why this should be the case we first observe that,
as we move into the domain from the left-hand boundary, the WKB approximations $v_3$ and $v_2$  decay exponentially, while $v_1$
grows  exponentially.
Let us start by describing branch (C). We  proceed to construct an eigenfunction as follows:
The left-hand boundary
condition can be satisfied by adding a multiple of $v_3$ and $v_2$.
Since $v_2$ exhibits a stronger decay, it can be safely disregarded at the right-hand
boundary.
To satisfy the conditions at the right-hand boundary we need some more degrees of freedom. We can add a multiple of $v_1$ without affecting the left-hand boundary ---since $v_1$ is exponentially larger on the right than the left. We choose our normalisation of the eigenfunction to fix this multiple of $v_1$ to unity (choosing $\phi_1(y_C)=0$).
Let us now determine the form of the eigenfunction at the right-hand boundary, taking into account the Stokes switchings.
The solution $v_1$ will turn on a multiple of $v_2$ due to the turning point $y_C$. The Stokes multiplier is $-1$ (if we set $\phi_2(y_C)=0$).
There are also Stokes switchings associated with $y_A$ and $y_B$, but we can see from Fig.~\ref{fig:StokesA} that the functions switched on are exponentially subdominant up to the boundary (since the anti-Stokes line has not been crossed).
As a result, the eigenfunction takes the following form near the right-hand boundary:
\begin{align}
\label{eq:eigen_func_right_BC}
v =  a A_3 \ee^{ \phi_3/\vareps} - A_2 \ee^{ \phi_2/\vareps} +
A_1 \ee^{ \phi_1/\vareps},
\end{align}
where $\phi_1$ and $\phi_2$ are both zero at $y=y_C$ (the zero of $\phi_3$ can be chosen arbitrarily since changing it simply changes the coefficient $a$).
Subsequently, using Eq.~\eqref{eq:eigen_func_right_BC}, we can express the first- and second-order derivatives of $v$ with respect to $y$ near the right-hand boundary as:
\begin{subequations}
\begin{align}
v' &= \frac{1}{\vareps} \left(a \phi_3' A_3 \ee^{ \phi_3/\vareps} - A_2 \phi_2' \ee^{ \phi_2/\vareps} +
A_1 \phi_1'\ee^{ \phi_1/\vareps}\right) + \cdots,
\label{eq:leading_vp}
\\
v''&=  \frac{1}{\vareps^2} \left(a (\phi_3')^2 A_3 \ee^{ \phi_3/\vareps} - A_2 (\phi_2')^2\ee^{ \phi_2/\vareps} +
A_1 (\phi_1')^2\ee^{ \phi_1/\vareps}\right) + \cdots.
\label{eq:leading_vpp}
\end{align}
\end{subequations}
Now imposing \eqref{eq:renKdV_BC_1} at $y = \vareps L$, the
coefficient $a$ can be determined at leading-order as
\begin{align}
\label{eq:a_expan}
a =  \frac{ A_2 \phi_2' \ee^{ \phi_2/\vareps} -
A_1 \phi_1'\ee^{ \phi_1/\vareps}}{\phi_3' A_3 \ee^{ \phi_3/\vareps}},
\end{align}
which reduces Eq.~\eqref{eq:leading_vpp} at $y = \vareps L$ to
\begin{align}
v'' &=   \frac{1}{\vareps^2} \left( \frac{ A_2 \phi_2' \ee^{ \phi_2/\vareps} -
A_1 \phi_1'\ee^{ \phi_1/\vareps}}{\phi_3' A_3 \ee^{ \phi_3/\vareps}} %
(\phi_3')^2 A_3 \ee^{ \phi_3/\vareps} -A_2 (\phi_2')^2\ee^{ \phi_2/\vareps} +
A_1 (\phi_1')^2\ee^{ \phi_1/\vareps}\right) + \cdots\nonumber \\
& =  \frac{1}{\vareps^2} \left(- A_2 \phi_2'(\phi_2'-\phi_3')\ee^{ \phi_2/\vareps} +
A_1 \phi_1'(\phi_1'-\phi_3')\ee^{ \phi_1/\vareps}\right) + \cdots.
\label{eq:leading_vpp_reduced}
\end{align}
From the above expression, we observe that $v''=0$ when the exponential terms
in Eq.~\eqref{eq:leading_vpp_reduced} are of comparable magnitude, requiring the following condition:
\begin{equation}
 \re\left(\phi_2 + \vareps \log \left( A_2 \phi_2' (\phi_2'-\phi_3')
    \right) \right) = \re\left(\phi_1 + \vareps \log \left( A_1 \phi_1' (\phi_1'-\phi_3')
    \right) \right),\label{answer1}
\end{equation}
so that the anti-Stokes line from $y_C$ must pass through the boundary $y = \vareps L$
(to leading order--equation (\ref{answer1}) includes an $O(\vareps)$ perturbation).

Setting (\ref{eq:leading_vpp_reduced}) to zero produces
complex eigenvalues with negative imaginary part and real part
around $G$.
If we instead require that the anti-Stokes line from
$y_A$ passes through $\vareps L$, then we produce the complex conjugate
of these eigenvalues, characterized by a positive imaginary part and real part
around $G$.

\begin{figure}[pt]
\begin{center}
\subfigure[]{
\begin{overpic}[width=0.45\textwidth]{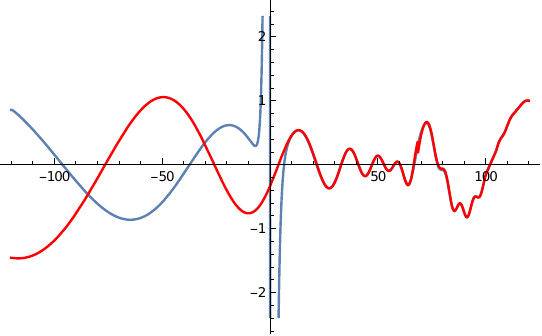}
\put(34,57){$\re(v)$}
\put(95,26){$\xi$}
\end{overpic}}
\subfigure[]{
\begin{overpic}[width=0.45\textwidth]{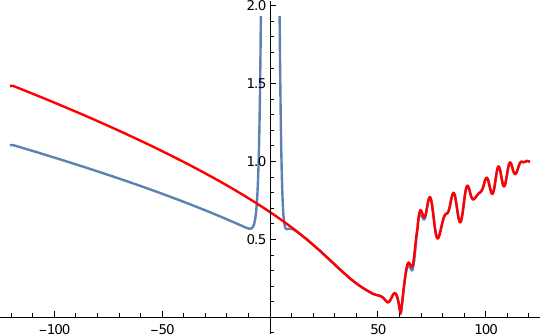}
\put(40,55){$|v|$}
\put(95,-3){$\xi$}
\end{overpic}}
\end{center}
\caption{The (a) real part and (b) absolute value of the eigenfunction
with eigenvalue  $\la \approx  -0.0243 - 0.1294\ii$ and
$G=-0.022$, $L=120$. The numerical solution
is shown in blue, the asymptotic approximation  in red. The asymptotic approximation
does not take into account the change in amplitude and phase of the WKB approximations
across the inner region near the origin, but this does not affect the eigenvalue or
the approximation of the eigenfunction in $y>0$.}
\label{fig:eigenfunction}
\end{figure}

Note that the  calculation above does not consider any potential alterations in the coefficients of the WKB solutions as we pass through the inner
region near $y=0$.
Nevertheless, since the key to determining the
eigenvalue lies in understanding the behavior on the right side of the turning points, which lie in the right half-plane,  and in the vicinity of  the right-hand boundary, any adjustments in these coefficients do not impact the determination of the eigenvalue.
This conclusion is evident from the representation presented in Fig.~\ref{fig:eigenfunction}, where we illustrate the asymptotic and numerical approximation to one eigenfunction. The asymptotic calculation predicts the eigenfunction well for $y>0$, but does not take into account the jump in the coefficients of the WKB solution as we pass through the inner region near $y=0$; nevertheless this does not affect the eigenvalue calculation.

\begin{figure}[pt]
\begin{center}
\begin{overpic}[width=0.6\textwidth]{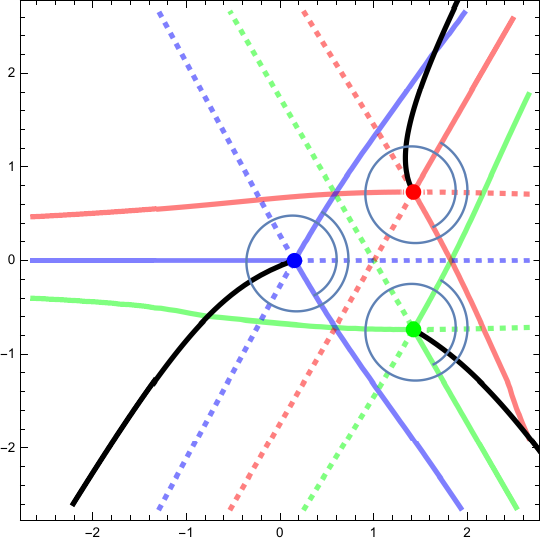}
\put(72,61){$y_A$}
\put(50,54){$y_B$}
\put(72,40){$y_C$}

\put(65,60){\small $\phi_1$}
\put(80,61){\small $\phi_3$}
\put(86,65){\small $\phi_1$}

\put(42,54){\small $\phi_3$}
\put(57,49){\small $\phi_3$}
\put(64,48){\small $\phi_2$}

\put(64,40){\small $\phi_1$}
\put(79,41){\small $\phi_1$}
\put(85,43){\small $\phi_2$}

\put(10,53){\small Stokes line}
\put(27,93){\rotatebox{-65}{\small Anti-Stokes line}}
\put(11,9){\rotatebox{58}{\small branch cut}}

\put(50,-5){$\re(y)$}
\put(-5,50){\rotatebox{90}{$\im(y)$}}
  \end{overpic}
\end{center}
\caption{Stokes lines (solid) and anti-Stokes lines (dashed) when $\la \approx -0.3$. Solid black lines are
branch-cuts of the phase function $\phi$. Three turning points $y_A$,
$y_B$ and $y_C$ are indicated. Each is a square root branch point of
$\phi$. The local Riemann surface structure in the vicinity of each
branch point is indicated by the spirals, which shows which branches
of $\phi$ are turned on/off across the Stokes lines.}
\label{fig:StokesB}
\end{figure}
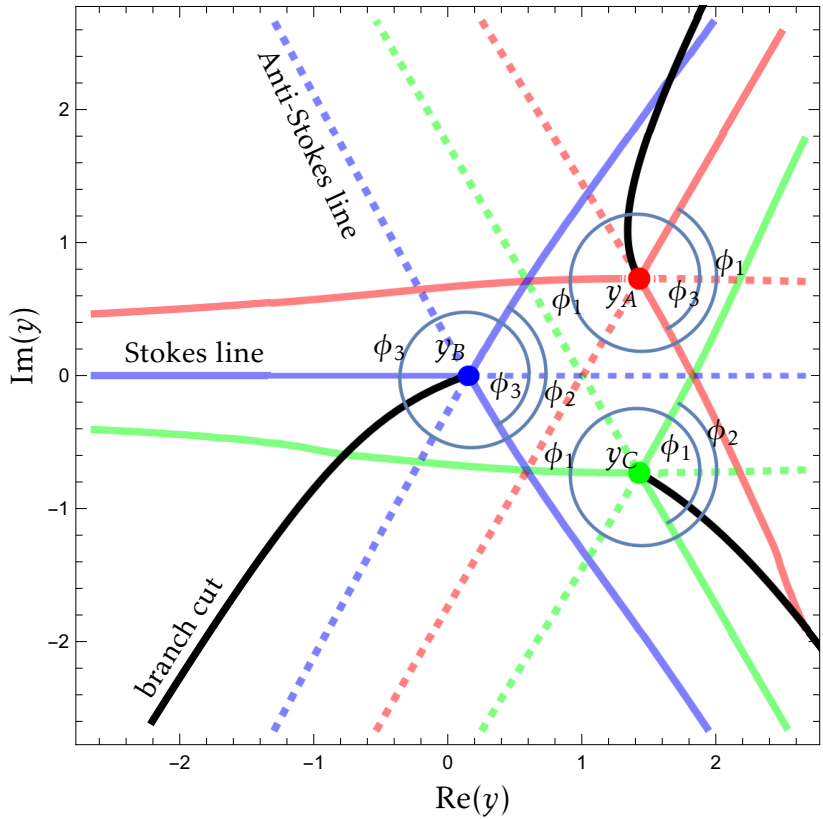

There is one final scenario that we need to consider, namely when the anti-Stokes
line originating from $y_B$  intersects the right-hand boundary $y = \vareps L$.
This occurs when the eigenvalue, $\la$, is real, with
the corresponding Stokes
line configuration  illustrated in Fig. \ref{fig:StokesB}.
In this configuration, the phases $\phi_3$ and
$\phi_2$ are approximately of the same magnitude along the real axis, leading to the expression:
\begin{align}
\label{eq:eigen_func_scenario_B}
v =   A_3 \ee^{ \phi_3/\vareps} - A_2 \ee^{ \phi_2/\vareps} + a A_1 \ee^{ \phi_1/\vareps}.
\end{align}
Similar to the previous analysis, we derive the following leading-order expressions for the derivative of $w$ with respect to $y$:
\begin{subequations}
\begin{align}
v' &=   \frac{1}{\vareps} \left( \phi_3' A_3 \ee^{ \phi_3/\vareps} - A_2 \phi_2' \ee^{ \phi_2/\vareps} +a
A_1 \phi_1'\ee^{ \phi_1/\vareps}\right)+\cdots,
\label{eq:leading_wp}
\\
v'' &=  \frac{1}{\vareps^2} \left(  (\phi_3')^2 A_3 \ee^{ \phi_3/\vareps} - A_2 (\phi_2')^2\ee^{ \phi_2/\vareps} +a
A_1 (\phi_1')^2\ee^{ \phi_1/\vareps}\right)+\cdots.
\label{eq:leading_wpp}
\end{align}
\end{subequations}
Imposing \eqref{eq:renKdV_BC_1} at $y = \vareps L$ we determine the coefficient $a$ to leading-order as
\begin{align}
a = - \frac{ A_3 \phi_3' \ee^{ \phi_3/\vareps} -
A_2 \phi_2'\ee^{ \phi_2/\vareps}}{\phi_1' A_1 \ee^{ \phi_1/\vareps}}.
\end{align}

\begin{figure}[pt]
\begin{center}
\subfigure[]{
  \begin{overpic}[width=0.45\textwidth]{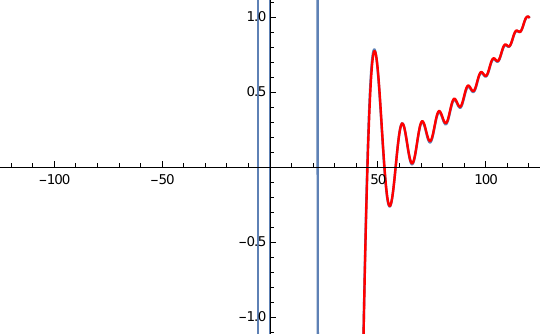}
\put(32,57){$\re(v)$}
\put(95,26){$\xi$}
  \end{overpic}}
\subfigure[]{
  \begin{overpic}[width=0.45\textwidth]{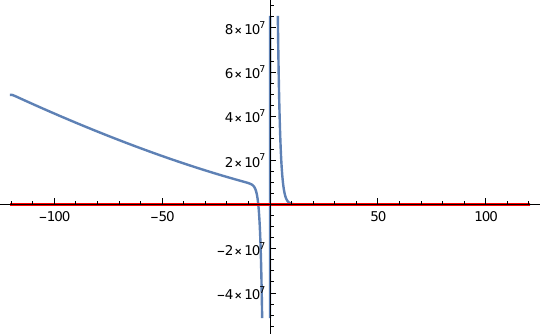}
\put(35,55){$|v|$}
\put(95,18){$\xi$}
  \end{overpic}}
\end{center}
\caption{The (a) real part and (b) absolute value of the eigenfunction
  with eigenvalue $\la \approx -0.0434$ for $G=-0.022$, $L=120$. The
  numerical solution
is shown in blue, and the asymptotic approximation in red.
Again the approximation does not take into account a jump in the
coefficient of the WKB solution near the origin, but this does not
affect the eigenfunction for positive $y$.}
\label{fig:eigenfunctionreal}
\end{figure}
This way, Eq.~\eqref{eq:leading_wpp} simplifies to:
\begin{align}
v'' &=  \frac{1}{\vareps^2}\left( A_3 (\phi_3')^2\ee^{ \phi_3/\vareps} -
A_2 (\phi_2')^2\ee^{ \phi_2/\vareps} - \frac{ A_3 \phi_3' \ee^{ \phi_3/\vareps} -
A_2 \phi_2'\ee^{ \phi_2/\vareps}}{\phi_1' A_1 \ee^{ \phi_1/\vareps}} %
(\phi_1')^2 A_1 \ee^{ \phi_1/\vareps}\right)+\cdots \nonumber \\
& =  \frac{1}{\vareps^2}\left(A_3 \phi_3'(\phi_3'-\phi_1')\ee^{ \phi_3/\vareps} -
A_2 \phi_2'(\phi_2'-\phi_1')\ee^{ \phi_2/\vareps}\right)+\cdots.
\end{align}
As previously, in order for this expression to be zero, the two exponentials must be
of comparable size.
Specifically, we derive:
\beq
\re\left(\phi_3 + \vareps \log \left( A_3 \phi_3' (\phi_3'-\phi_1')
\right) \right) = \re\left(\phi_2 + \vareps \log \left( A_2 \phi_2' (\phi_2'-\phi_1')
\right) \right).
\label{real1}
\eeq
When $\la$ is real, for real $y$,  $\phi_1'$ and $y_{B}$ are real, and $\phi_2 =
\bar{\phi}_3$ (with the overbar standing for complex conjugation).
Consequently, $w''(\vareps L)=0$ if
\beqas
\ee^{ 2 \ii \im(\phi_3)/\vareps} =
\frac{\bar{\phi}_3'(\bar{\phi}_3'-\phi_1')}{(3
  (\bar{\phi}_3')^2-1+\vareps L)^{1/2}}\frac{(3
  ({\phi}_3')^2-1+\vareps L)^{1/2}}{\phi_3'(\phi_3'-\phi_1')} = \ee^{
  \ii \Phi},
\eeqas
say, i.e.
\begin{equation}
  \im(\phi_3) = \vareps\left(n \pi  + \frac{ \Phi }{2}\right).
  \label{realeigsouter}
\end{equation}
An illustrative example is presented in Fig.~\ref{fig:eigenfunctionreal}, which compares the numerically obtained eigenfuction with the asymptotic
approximation for a real eigenvalue $\lambda\approx -0.0434$.

\begin{figure}[pt]
\begin{center}
\begin{overpic}[width =
  0.75\textwidth]{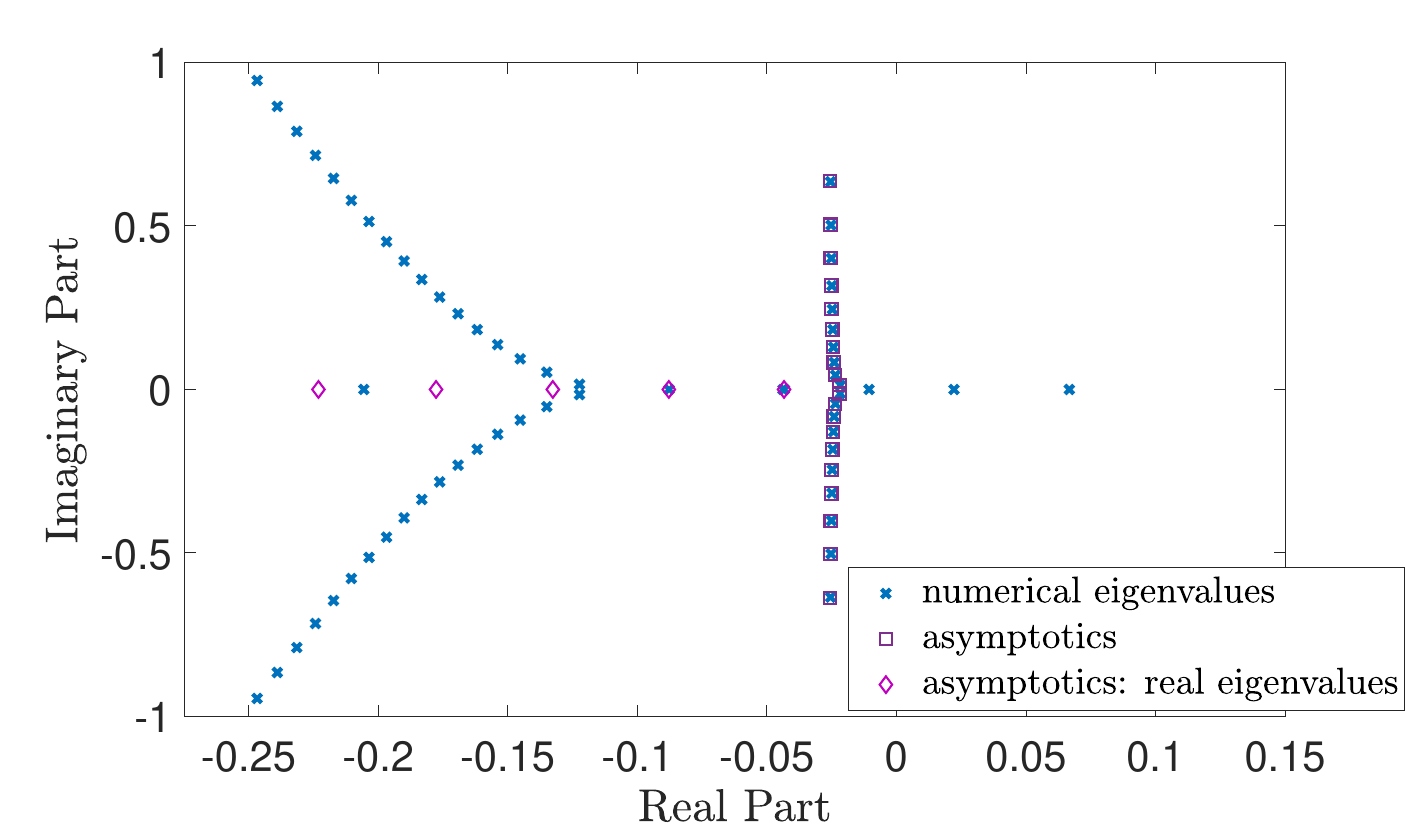}
  \put(59,40){(A)}
\put(59,22){(C)}
\put(44,33){(B)}

\end{overpic}
\caption{Comparison between asymptotic approximation (rectangles and
  diamonds) and numerical eigenvalues (blue crosses) for domain
  length, $L=120$, and parameter, $p=5.1$ ($G \approx
  -0.022$). Labels (A), (B) and (C) correspond to the different
  branches of the asymptotic approximation.
\label{fig:numerics_vs_asmyptotic}}
\end{center}
\end{figure}

In Figure \ref{fig:numerics_vs_asmyptotic} we show a comparison of the
asymptotic predictions for the eigenvalues along with the
numerically-determined spectrum for $p=5.1$ and a domain of length
$L=120$. We see that there is excellent agreement for the complex
eigenvalues in branches (A) and (C), and for the first few real
eigenvalues in branch (B). However, moving to the left the numerically
determined eigenvalues do not stay on the real line,  but diverge into
a wedge in the complex plane. We believe that this is numerical error,
and is caused by the extreme ill-conditioning of the
eigenvalue problem. The origin of this can be seen in Figure
\ref{fig:eigenfunctionreal}---these eigenfunctions are exponentially
small near the
right-hand boundary by comparison to the left-hand boundary (a factor
of $10^7$ in Fig.~\ref{fig:eigenfunctionreal} where $\la = -0.0434$
and $G=-0.022$), and yet the
eigenvalue condition is determined by the behaviour at the right-hand
boundary.

The previous calculation of real eigenvalues disregards Stokes switching due to the turning
points at $y_A$ and $y_C$, which  remain exponentially subdominant until $\la$ reaches the region in the complex plane where the first two
branches of eigenvalues (A) and (C)  intersect the real axis, i.e. around $\la \approx G$. Thus the approximations we have found are good at describing each branch of eigenvalues away from the point of intersection of the branches.
To get a good approximation near the intersection point we need to incorporate the effects of all the Stokes lines. Doing so  we arrive  at the following expression
\begin{align}
v &=   A_3 \ee^{ \phi_3/\vareps} - A_2 \ee^{ \phi_2/\vareps}
  + a
\left(A_1 \ee^{ \phi_1/\vareps} -A_3 \ee^{
(\phi_1(y_A)-\phi_3(y_A))/\vareps}\ee^{\phi_3/\vareps} - A_2
    \ee^{ (\phi_1(y_C)-\phi_2(y_C))/\vareps}\ee^{
      \phi_2/\vareps}\right)  \non \\
   &=  (1-a \ee^{
      (\phi_1(y_A)-\phi_3(y_A))/\vareps}) A_3 \ee^{ \phi_3/\vareps} -
    (1+ a \ee^{ (\phi_1(y_C)-\phi_2(y_C))/\vareps})A_2 \ee^{ \phi_2/\vareps} + a
 A_1 \ee^{ \phi_1/\vareps}.
\label{efunclose}
\end{align}
Differentiating Eq.~\eqref{efunclose} with respect to $y$,
and considering the leading order, we obtain:
\begin{subequations}
\begin{align}
v'&=\frac{1}{\vareps}\left((1-a \ee^{
      (\phi_1(y_A)-\phi_3(y_A))/\vareps})\phi_3' A_3 \ee^{
      \phi_3/\vareps} -  (1+ a
      \ee^{(\phi_1(y_C)-\phi_2(y_C))/\vareps}) A_2 \phi_2' \ee^{ \phi_2/\vareps}+aA_1 \phi_1'\ee^{ \phi_1/\vareps}\right)+\cdots,
  \label{eq:leading_new_wp}
  \\
v''&=\frac{1}{\vareps^2}\left((1-a \ee^{
      (\phi_1(y_A)-\phi_3(y_A))/\vareps}) (\phi_3')^2 A_3 \ee^{ \phi_3/\vareps} - (1+ a \ee^{ (\phi_1(y_C)-\phi_2(y_C))/\vareps}) A_2 (\phi_2')^2\ee^{ \phi_2/\vareps}
      \non \right. \\ & \left. \hspace{11cm}+aA_1 (\phi_1')^2\ee^{
        \phi_1/\vareps} \right)+\cdots.
  \label{eq:leading_new_wpp}
\end{align}
\end{subequations}
The value of $a$ in Eq.~\eqref{eq:leading_new_wp} can be determined
using \eqref{eq:renKdV_BC_1} as usual, giving
\begin{align}
\label{eq:leadning_new_a}
a = - \frac{ A_3 \phi_3' \ee^{ \phi_3/\vareps} -
A_2 \phi_2'\ee^{ \phi_2/\vareps}}{\phi_1' A_1 \ee^{ \phi_1/\vareps}- \ee^{
      (\phi_1(y_A)-\phi_3(y_A))/\vareps}\phi_3' A_3 \ee^{
      \phi_3/\vareps}  -  \ee^{
      (\phi_1(y_C)-\phi_2(y_C))/\vareps} A_2 \phi_2' \ee^{ \phi_2/\vareps} }.
\end{align}
By substituting Eq.~\eqref{eq:leadning_new_a} into Eq.~\eqref{eq:leading_new_wpp},
we derive:
\begin{multline*}
  v'' =
  -\frac{1}{\vareps^2}\frac{A_3 \phi_3' \ee^{ \phi_3/\vareps}-A_2 \phi_2'\ee^{ \phi_2/\vareps}}%
{\phi_1' A_1 \ee^{ \phi_1/\vareps}%
-\ee^{(\phi_1(y_A)-\phi_3(y_A))/\vareps}\phi_3' %
A_3 \ee^{\phi_3/\vareps}-\ee^{(\phi_1(y_C)-\phi_2(y_C))/\vareps} %
A_2 \phi_2' \ee^{\phi_2/\vareps}}\\
\times\Bigg\lbrace
(\phi_1')^2 A_1\ee^{\phi_1/\vareps}%
-\ee^{(\phi_1(y_A)-\phi_3(y_A))/\vareps}(\phi_3')^2 A_3 %
\ee^{\phi_3/\vareps}%
-\ee^{(\phi_1(y_C)-\phi_2(y_C))/\vareps} A_2 (\phi_2')^2 \ee^{ \phi_2/\vareps}
\Bigg\rbrace \\
+\frac{1}{\vareps^2}\left(A_3 (\phi_3')^2\ee^{ \phi_3/\vareps} -A_2 (\phi_2')^2\ee^{ \phi_2/\vareps}\right)+\cdots.
\end{multline*}
Consequently, after rearranging terms, the eigenvalue condition can be expressed as
\begin{multline}
\frac{A_3 (\phi_3')^2\ee^{ \phi_3/\vareps} -
A_2 (\phi_2')^2\ee^{ \phi_2/\vareps}}{A_3 \phi_3' \ee^{ \phi_3/\vareps} -
A_2 \phi_2'\ee^{ \phi_2/\vareps}} \\
=\frac{(\phi_1')^2 A_1 \ee^{
\phi_1/\vareps} -  \ee^{
(\phi_1(y_A)-\phi_3(y_A))/\vareps}(\phi_3')^2 A_3 \ee^{
\phi_3/\vareps}  -  \ee^{
(\phi_1(y_C)-\phi_2(y_C))/\vareps} A_2 (\phi_2')^2 \ee^{ \phi_2/\vareps}}{\phi_1' A_1 \ee^{ \phi_1/\vareps}
- \ee^{(\phi_1(y_A)-\phi_3(y_A))/\vareps}\phi_3' A_3
\ee^{\phi_3/\vareps}
- \ee^{(\phi_1(y_C)-\phi_2(y_C))/\vareps} A_2 \phi_2' \ee^{\phi_2/\vareps}}.
\label{evcondition}
\end{multline}

\begin{figure}
  \begin{center}
    \begin{overpic}[width=0.6\textwidth]{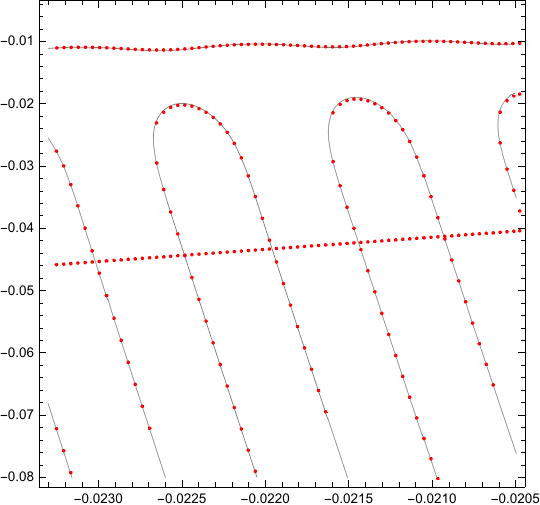}
      \put(50,-4){$G$}
      \put(-4,50){$\la$}
      \end{overpic}
\end{center}
\caption{
Negative real eigenvalues of the
operator $\mathcal{L}$ given by Eq.~\eqref{eq:renKdV_eval_prob_op},
branch (B),  near the intersection with branches (A) and (C).
The real eigenvalues as a function of $G$
are shown with red dots and solid grey line corresponding
to the numerical solution of the eigenvalue problem [cf. Eq.~\eqref{eq:renKdV_eval_prob}]
and asymptotic approximation of Eq.~\eqref{evcondition},
respectively. Note that although the formula captures the
slight oscillation of the eigenvalue located near $G/2$,
there is a discrete eigenvalue near $2G$ which is missing.}
\label{fig:eigenvaluesVsG}
\end{figure}

In Fig.~\ref{fig:eigenvaluesVsG}, we illustrate the accuracy of (\ref{evcondition}) in predicting the complicated behaviour of the negative real eigenvalues  associated with the spectrum of
self-similar solutions to the gKdV equation as $G$ is varied.
The red dots in the plot correspond to the eigenvalues obtained through numerical computation, as visualized previously in Fig.~\ref{fig:dominanteigs_negative}.
The solid gray line, on the other hand, represents the asymptotic approximation provided by Eq.~\eqref{evcondition}.
The asymptotic formula accurately captures the vast majority of negative eigenvalues, with the exception of a discrete eigenvalue located at $2G$.

Ignoring for the moment the eigenvalue near $\la\approx G/2$,
as $G$ is reduced all other negative eigenvalues move towards the origin. The leading eigenvalue is then ``reflected" near $\la\approx G$ and returns to collide with its neighbour (at which point they become complex and add to the eigenvalue branches (A) and (C)).

This eigenvalue dance induces a subtle oscillatory behavior in the eigenvalue near $G/2$ (that was also illustrated in the right panel of
Fig.~\ref{fig:dominanteigs_negative}, indicated by the solid line and blue open circles), which is also captured well by Eq.~\eqref{evcondition}.

\begin{figure}
\begin{center}
\begin{overpic}[width=0.45\textwidth]{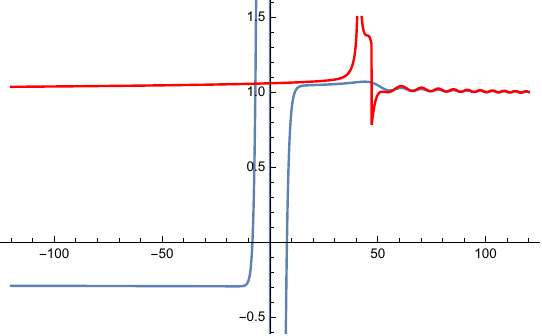}
\put(32,57){$\re(v)$}
\put(95,10){$\xi$}
\end{overpic}
\end{center}
\caption{The numerically obtained and
asymptotically constructed (from Eq.~\eqref{efunclose})
eigenfunction with eigenvalue $\la\approx G/2$ is shown with
a solid blue and red lines, respectively. The outer solution
breaks down near the turning point at $y_B$, but captures well
the oscillations to the right of this.}
\label{fig:eigenfunctionisolatedALT}
\end{figure}

Finally we note that the leading-order outer equation [cf. Eq.~\eqref{eq:leading}]
and boundary conditions are satisfied by $v \equiv 1$ when
$\la = - \vareps/2$, which explains why there is an eigenvalue near $\la \approx G/2$.
However, the eigenfunction predicted by Eq.~\eqref{efunclose} is not quite constant in $y>0$,
as illustrated by the solid red line in Fig.~\ref{fig:eigenfunctionisolatedALT}; the approximation
effectively captures the small oscillations present in the eigenfunction to the right of
the turning point at $y_{B}$.

\subsection{The normal form}\label{sec_3_4}

After delving into the spectrum of self-similar solutions in Sec.~\ref{sec_3_3}, our focus now shifts towards the derivation of the normal form, a pivotal model equation characterizing the emergent bifurcation of the self-similar branch from from the soliton branch at $p=5$.
The final expression of this normal form is provided in Eq.~\eqref{normalform} for convenient reference.
In the previous section we defined $\vareps=-G$, expanding $p = p_0 + \vareps p_1 + \cdots$.
In this section we find it convenient to set  $p = p_0 + \vareps p_1$, and expand $G$ in powers of $\eps$. We could of course set $p_1=1$, but since we would like to maintain $\vareps>0$ while considering both $p>p_0$ and $p<p_0$ we find it more convenient to leave $p_1$ arbitrary.
This approach also helps us clearly identify the origin of the terms within the expansion.

We  transition to a slower time scale by introducing the transformation
$t = \vareps \tau$. Then Eq.~\eqref{eq:renKdV_v2} takes the form
\begin{align}
\label{1b}
\varepsilon \pd{w}{t} = - \frac{\dd^3 w}{\dd\xi^3} %
- \pd{w^p}{\xi} + G\left(\frac{2}{p-1} w + \xi \pd{w}{\xi} \right) %
+ \pd{w}{\xi},
\end{align}
and we note that the blow-up rate, $G$, continues to be uniquely determined by the
pinning condition described in Eq.~\eqref{pinning_cond}.
Much of the analysis in this section mirrors that in
\cite{weinmueller_2020} for the
steady problem.
We expand the solution $w$ and the (constant) blow-up rate $G$ in powers of $\vareps$ as
\begin{align}
\label{eq:power_series}
w = \sum_{n=0}^\infty \vareps^n w_n,  \qquad G = \sum_{n=1}^\infty
\vareps^n G_n.
\end{align}
Substituting Eq.~\eqref{eq:power_series} into Eq.~\eqref{1b},
we obtain at order $\mathcal{O}(1)$:
\begin{align}
\frac{\d^3 w_0}{\d\xi^3} + \frac{\d}{\d \xi}\left({w_0^{p_0}} \right) - \frac{\d w_0}{\d \xi}=0.
\end{align}
This equation is just that describing steady travelling waves to Eq.~\eqref{eq:genKdV}, with  solution [cf.~Eq.~\eqref{eq:soliton}]
\begin{align}
\label{lead_in}
w_{0}=\left(\frac{p_{0}+1}{2}\right)^{\frac{1}{p_{0}-1}}%
\sech^{\frac{2}{p_{0}-1}}{\left(\frac{\left(p_{0}-1\right)}{2}%
\left(\xi-\xi_{0}\right)\right)}.
\end{align}
Note that the offset $\xi_0$ in Eq.~\eqref{lead_in} is determined by the pinning condition specified in Eq.~\eqref{pinning_cond}, which in this
case results in $\xi_0=0$.

At order $\mathcal{O}(\vareps)$, considering that $\pd{w_0}{t}=0$ and integrating
once with respect to $\xi$, we find:
\begin{align}
{\cal L}^{(1)} w_1 &=%
- p_1 w_0^{p_0}\log w_0 +G_1 \int_{0}^\xi\left(\frac{2}{{p_0}-1}w_0%
+ s \pd{w_0}{s} \right) \, \d s + c \nonumber \\
& =  - p_1 w_0^{p_0}\log w_0 +G_1 g+ c,\label{w1eqn}
\end{align}
where
\begin{align}
{\cal L}^{(1)}\left(\cdot \right)\coloneqq %
\frac{\dd^2}{\dd\xi^2} + p_0 w_0^{p_0-1} - 1, \quad %
g=\int_{0}^\xi\left(\frac{2}{{p_0}-1} w_0 + s \pd{w_0}{s} \right) \, \d s,
\label{eq:ops}
\end{align}
and $c$ is an arbitrary constant of integration.
Note that since
$w_0$ [cf.~Eq.~\eqref{lead_in}] is an even function, $g$ [cf.~Eq.~\eqref{eq:ops}]
is odd.
Additionally, we note that since $\pd{w_0}{\xi}$ satisfies the
homogeneous version of Eq.~\eqref{w1eqn}, a solvability condition arises due to the Fredholm alternative.
This condition is expressed as follows:
{\small
\begin{align}
0 & = G_1\int_{-\infty}^\infty \pd{w_0}{\xi}%
\int_{0}^\xi\left(\frac{2}{{p_0}-1} w_{0} + s\pd{w_{0}}{s} \right)\, \d s \, \d \xi \nonumber \\
&=  -G_1\int_{-\infty}^\infty w_0
\left(\frac{2}{{p_0}-1} w_0 + \xi \pd{w_0}{\xi} \right) \, \d \xi \nonumber \\
& =
-G_1\int_{-\infty}^\infty
\left(\frac{2}{{p_0}-1} w_0^2 - \frac{w_0^2}{2} \right) \, \d \xi=
\frac{G_1({p_0}-5)}{2({p_0}-1)}\int_{-\infty}^\infty w_0^2 \, \d \xi,
\label{eq:solvability_cond_1}
\end{align}
}
due to the relations
\begin{align*}
\int_{-\infty}^\infty \pd{w_0}{\xi} w_0^{p_0} \log w_0 \, \d \xi = 0, %
\qquad \int_{-\infty}^\infty \pd{w_0}{\xi}  \, \d \xi = 0.
\end{align*}

From Eq.~\eqref{eq:solvability_cond_1}, it follows that $p_0=5$ when $G_1 \not = 0$~\cite{jon_expl_1}.
Furthermore, the complementary functions can be expressed as:
\begin{align*}
v_1 = \frac{\sinh(2 \xi)}{\cosh^{3/2}(2 \xi)}, \qquad %
v_2 = \frac{\cosh(4 \xi) - 3}{\cosh^{3/2}(2 \xi)},
\end{align*}
and they satisfy
\begin{align*}
v_1(0) = 0, \quad v_2'(0) = 0, \qquad v_1 \ra 0, \quad %
v_2 \ra \infty \mbox{ as } |\xi| \ra \infty
\end{align*}
while also having a Wronskian
\begin{align*}
W = v_1 v_2' - v_2 v_1' = 4.
\end{align*}
As a result, we can express $w_1$ as:
\begin{equation}
w_1 = v_1 \int_0^\xi \frac{v_2}{4}\left( p_1 w_0^{p_0}\log w_0
  -G_1 g-c\right) \, \d s
  + v_2 \int_\xi^\infty
\frac{v_1}{4}\left(p_1 w_0^{p_0}\log w_0 -G_1 g - c\right)\, \d s,
\end{equation}
with reference to ~\cite{jon_expl_2}
{\small
\begin{subequations}
\begin{align}
w_1&\sim-c - G_1g(\infty) = -c + \frac{G_1}{2} \int_0^\infty
w_0(\xi)\, \d \xi = -c +\frac{3^{1/4} \Gamma(1/4)^2G_1 }{8 \sqrt{2 \pi}} %
\quad \mbox{as } \xi \ra \infty,\label{in_inf}\\
w_1 & \sim-c -G_1 g(-\infty) = -c -\frac{ 3^{1/4} \Gamma(1/4)^2G_1}{8 \sqrt{2 \pi}}  %
\quad \mbox{as } \xi \ra- \infty.\label{in_minf}
\end{align}
\end{subequations}
}
To determine the constant, $c$, we delve into a more detailed analysis of the far-field region, as discussed in Sec.~\ref{outer}. First, though, we proceed to one more order in this inner expansion.

At order $\mathcal{O}\left(\vareps^{2}\right)$, upon
integrating once with respect to $\xi$ and rearranging,
the result is:
\begin{align*}
{\mathcal L}^{(1)} w_2 & = - \int_{-\infty}^\xi\pd{w_1}{t}\,\d s %
- 10 w_0^3 w_1^2 - p_1 w_0^4 w_1 - 5 p_1 w_0^4 w_1 \log w_0 %
- \frac{p_1^2}{2} w_0^5 \log^2 w_0\nonumber \\
&+G_1\int_{-\infty}^\xi \left(\frac{w_1}{2} + s\pd{w_1}{s} %
- \frac{p_1}{8} w_0 \right)\, \d s%
+c_1+G_2 \int_{0}^\xi\left(\frac{w_0}{2}+ s \pd{w_0}{s} \right) \, \d s,
\end{align*}
where $c_{1}$ is an arbitrary constant of integration,
%
%

In this case, the solvability condition is given by
\begin{align*}
0 & =  \int_{-\infty}^\infty \pd{w_0}{\xi}\left(-  \int_{-\infty}^\xi
\pd{w_1}{t}\, \d s- 10 w_0^3 w_1^2 - p_1 w_0^4 w_1 %
- 5 p_1 w_0^4 w_1 \log w_0\right.\\
& \qquad \qquad\left.\mbox{  } - \frac{p_1^2}{2} w_0^5 \log^2 w_0%
+G_1\int_{-\infty}^\xi \left(\frac{w_1}{2} + s \pd{w_1}{s} %
- \frac{p_1}{8} w_0 \right)\, \d s\right)\, \d \xi \\
& = \int_{-\infty}^\infty \pd{w_0}{\xi}%
\left(- 10 w_0^3 w_1^2 - p_1 w_0^4 w_1 - 5 p_1 w_0^4 w_1 \log w_0\right)\, \d \xi\\&
 \qquad\qquad \mbox{  }
+G_1\int_{-\infty}^\infty w_0 \left(\frac{w_1}{2}  + \frac{p_1}{8} w_0 \right)\, \d \xi
+ w_0 \pd{w_1}{t}+G_1\int_{-\infty}^\infty  \xi\pd{w_0}{\xi}w_1\, \d \xi.
\end{align*}
Note that there is a $p_1$ term in $w_1$, so this equation is actually
quadratic in $p_1$.
We can explicitly express the dependence of of $w_1$ on $p_1$, $G_1$, and $c$ by writing $w_1 = p_1 W_1 +  G_1
\hat{w}_{\mathrm{odd}} + c
\hat{w}_{\mathrm{even}}$ where
\begin{align*}
W_1 & = v_1 \int_0^\xi \frac{v_2}{4}\left(  w_0^{5}\log w_0\right) \, \d s + v_2 \int_\xi^\infty \frac{v_1}{4}\left(  w_0^{5}\log w_0\right)\, \d s,\\
\hat{w}_{\mathrm{odd}} & =  -v_1 \int_0^\xi \frac{v_2 g}{4}  \, \d
s - v_2 \int_\xi^\infty \frac{v_1 g}{4} \, \d s,\\
\hat{w}_{\mathrm{even}} & =  -v_1 \int_0^\xi \frac{v_2}{4} \, \d s %
- v_2 \int_\xi^\infty \frac{v_1}{4}\, \d s.
\end{align*}
The only time dependence in $w_1$ arises from $G_1$ and $c$.
Consequently, the solvability condition simplifies to
\begin{multline}
G_1\int_{-\infty}^\infty \pd{w_0}{\xi}\left(- 10 w_0^3 2(p_1 W_1  +c \hat{w}_{\mathrm{even}} )\hat{w}_{\mathrm{odd}} - p_1 w_0^4 ( \hat{w}_{\mathrm{odd}} )
 - 5 p_1 w_0^4 (\hat{w}_{\mathrm{odd}}  )
 \log w_0\right)\, \d \xi\\
+\fdd{c}{t} \int_{-\infty}^\infty w_0
\hat{w}_{\mathrm{even}}\, \d \xi
+G_1\int_{-\infty}^\infty w_0 \left(\frac{(p_1 W_1  +c \hat{w}_{\mathrm{even}} )
}{2}  + \frac{p_1}{8} w_0 \right)\, \d \xi \\
 \mbox{ }
+G_1\int_{-\infty}^\infty  \xi\pd{w_0}{\xi}(p_1 W_1  +c \hat{w}_{\mathrm{even}} )
\, \d \xi = 0.
\end{multline}

Notably, all quadratic terms in $p_1$
cancel out (the Authors in~\cite{weinmueller_2020} get away
with omitting these terms).
Refining the solvability condition, we arrive at
\beq
  \alpha \fdd{c}{t} = G_1( \beta p_1 + \gamma c),\label{norm}
\eeq
where
\begin{align*}
\alpha & = -\int_{-\infty}^\infty w_0
\hat{w}_{\mathrm{even}}\, \d \xi, \\
\beta & =  \int_{-\infty}^\infty \pd{w_0}{\xi}%
\left(- 20 w_0^3  W_1   -  w_0^4 - 5  w_0^4 \log w_0\right)\hat{w}_{\mathrm{odd}}%
+ w_0 \left(\frac{ W_1}{2}  + \frac{1}{8} w_0 \right)+  \xi\pd{w_0}{\xi} W_1
\, \d \xi,\\
\gamma & =   \int_{-\infty}^\infty \left( -20\pd{w_0}{\xi}  w_0^3   \hat{w}_{\mathrm{odd}}
+ \frac{w_0}{2}
+  \xi\pd{w_0}{\xi}\right) \hat{w}_{\mathrm{even}}
\, \d \xi.
\end{align*}
Numerically, the coefficients evaluate to:
\[
\alpha  \approx  0.862705,\qquad
\beta  \approx  0.340087,\qquad
\gamma  \approx 1.72541.
\]
Thus, the steady state~\cite{jon_expl_3} of Eq.~\eqref{norm}
becomes
\begin{align*}
p_1 = -c\frac{\gamma}{\beta}  \approx  -5.07 c.
\end{align*}

It remains to determine $c$. To do so we need to consider the outer region in which $\xi$ is large.

\subsubsection{Outer region}\label{outer}
We now consider the far-field region where $|\xi|\gg 1$. In this case,
$w^p$ is negligible since $w$ is exponentially small. Writing $y = \vareps \xi$ as in
\eqref{eq:resc_outer} gives
\begin{equation}
\label{eq:renKdV_v3_far_field_2}
\vareps^2 \frac{\dd^3 w}{\dd y^3}-\pd{w}{y}  =
G_1\left(\frac{w}{2}  + y \pd{w}{y}\right) -\frac{\vareps G_1p_1 w }{8} %
+ \frac{\vareps^{2}p_{1}^{2}wG_{1}}{32}-\cdots.
\end{equation}
As in Sec.~\ref{sec_3_3}, WKB solutions are sought by
using the ansatz
\begin{align}
w \sim A \ee^{\phi/\vareps}, \quad \vareps \downarrow 0,
\end{align}
and this way, from Eq.~\eqref{eq:renKdV_v3_far_field_2},
we obtain:
\begin{subequations}
\begin{align}
(\phi')^3 - \phi' &= G_1 y \phi',
\label{eq1_WKB_2} \\
3 A' (\phi')^2 + 3 A \phi' \phi'' - A' &= G_1\frac{A}{2} +G_1 y A',
\label{eq2_WKB_2}
\end{align}
\end{subequations}
at orders $\mathcal{O}(\vareps^{-1})$ and $\mathcal{O}(1)$, respectively.
From Eq.~\eqref{eq1_WKB_2}, we find
\begin{align}
\label{eq:WKB_2_phip}
\phi'=\pm \left(1+G_{1}y\right)^{1/2},
\end{align}
after which
Eq.~\eqref{eq2_WKB_2} can be integrated to give
\begin{align}
\label{eq:WKB_2_A}
A=\mbox{const.} (1+G_1 y)^{-1/2}.
\end{align}
Since $\phi'=0$ is also a solution to \eqref{eq1_WKB_2} there is also a non-WKB solution. Setting $\vareps=0$ in \eqref{eq:renKdV_v3_far_field_2} we find that at leading order this solution is
\begin{align*}
w \sim \mbox{const.}  (1+G_1 y)^{-1/2}.
\end{align*}
Putting everything together, we have
\begin{align}
\label{eq:WKB_w_sol}
w \sim \frac{1}{{(1+G_1y)^{1/2}}}%
\left(a_1+a_2\ee^{-\frac{1}{\vareps}\int_0^y(1+G_1s)^{1/2}\, \d s}%
+a_3\ee^{\frac{1}{\vareps}\int_0^y(1+G_1s)^{1/2}\, \d s}\right),
\end{align}
for some constants $a_1$, $a_2$ and $a_3$.
Note the existence of a turning point at $y=-1/G_1$. This will be crucial in the analysis to follow.

We next proceed by separating the cases with $G_{1}<0$ and $G_{1}>0$.
We begin first with the case $G_{1}<0$.

\subsubsection{$G_1<0$}
For $y<0$ we need $a_2=0$ in Eq.~\eqref{eq:WKB_w_sol} otherwise we have
exponential growth. So, Eq.~\eqref{eq:WKB_w_sol} reduces to
\[ w \sim  \frac{a_1}{(1+G_1y)^{1/2}}
  + \frac{a_3}{(1+G_1y)^{1/2}}\ee^{\frac{2}{3G_1 \vareps}((1+G_1y)^{3/2}-1)}.
\]
Now, in terms of the inner variable, we have:
\[ w \sim  \frac{a_1}{(1+G_1 \vareps \xi)^{1/2}}
  + \frac{a_3}{(1+G_1\vareps
    \xi)^{1/2}}\ee^{\frac{2}{3G_1\vareps}((1+G_1 \vareps\xi)^{3/2}-1)}
\sim a_1 + a_3 \ee^{\xi} + \cdots,
\]
and matching with Eq.~\eqref{lead_in}, Eq.~\eqref{in_minf} gives
\begin{align*}
a_1  =\left( -c - \frac{3^{1/4}\Gamma(1/4)^2 G_1}{8 \sqrt{2 \pi}}\right)\vareps, \qquad
a_3  =12^{1/4}.
\end{align*}
On the other hand, and for $0<y<-1/G_1$, we have
\begin{align*}
w \sim \frac{1}{(1+G_1 y)^{1/2}}\left(b_1 +
  b_2\ee^{-\frac{2}{3G_1\vareps}((1+G_1 y)^{3/2}-1)}%
  +b_3\ee^{\frac{2}{3G_1\vareps}((1+G_1y)^{3/2}-1)}\right),
\end{align*}
for some new constants $b_i$. Matching with Eq.~\eqref{lead_in}, Eq.~\eqref{in_inf}
gives $b_3=0$, as well as
\begin{align*}
b_2 = 12^{1/4}, \qquad %
b_1 =\left(-c + \frac{3^{1/4}\Gamma(1/4)^2 G_1}{8 \sqrt{2 \pi}}\right)\vareps.
\end{align*}
For $y$ past the turning point, i.e., $y>-1/G_1$, the
WKB solution assumes the form
\begin{align*}
w \sim \frac{1}{(-G_1y-1)^{1/2}}\left(c_1+c_2\ee^{-\frac{2\ii}{3G_1\vareps}(-G_1y-1)^{3/2}}%
+c_3\ee^{\frac{2\ii}{3G_1\vareps}(-G_1y-1)^{3/2}}\right)
\end{align*}
for some new constants $c_i$.
The behaviour at infinity means that $c_2=c_3=0$. %

Finally, we need to connect the coefficients $b_i$ with $c_i$ by matching with a solution in the vicinity of the turning point. Near the turning point, we rescale $y$ in
Eq.~\eqref{eq:renKdV_v3_far_field_2} according to
$y = -1/G_1 + (2\vareps)^{2/3}z/G_1^{1/3}$, to give,
at leading order,
\begin{align*}
\frac{\dd^3 w}{\dd z^3}& =   2w  + 4z \pd{w}{z} ,
\end{align*}
with solution given in terms of Airy functions:
\[ w = \alpha_1 \Ai(z)^2 + \alpha_2 \Ai(z)\Bi(z) + \alpha_3 \Bi(z)^2.\]
Those have the asymptotic behaviors
\[ \Ai(z)^2 \sim \frac{1}{\pi |z|^{1/2}} \sin^2(2/3 |z|^{3/2}+\pi/4),\qquad
  \Bi(z)^2 \sim \frac{1}{\pi |z|^{1/2}} \cos^2(2/3 |z|^{3/2}+\pi/4)\]
as $z \ra -\infty$, and also
\[ \Ai(z)^2 \sim \frac{1}{4 \pi z^{1/2}} \ee^{-4/3 z^{3/2}},\qquad
  \Bi(z)^2 \sim \frac{1}{\pi z^{1/2}} \ee^{4/3 z^{3/2}},\]
as $z \ra \infty$.
The inner limit of the outer is
\beqas
w &=& \frac{b_1}{|2G_1|^{1/3}\vareps^{1/3}z^{1/2}} + \frac{b_2}{|2G_1|^{1/3}z^{1/2}\vareps^{1/3}}\ee^{2/3G_1\vareps}\ee^{4/3 z^{3/2}} \qquad z>0,\\
w &=& \frac{c_1}{|2G_1|^{1/3}(-z)^{1/2}\vareps^{1/3}}  \qquad z<0.
\eeqas
To match as $z \ra -\infty$ requires
\[\alpha_1=\alpha_3=\frac{c_1\pi }{|2G_1|^{1/3}\vareps^{1/3}}, \qquad \alpha_2=0,\]
whereas, to match as $z \ra \infty$ requires:
\[ \alpha_3 = \frac{b_2\pi}{|2G_1|^{1/3}\vareps^{1/3}}\ee^{2/3G_1\vareps}, \qquad b_1 = 0.
\]
Thus we finally see that
\[ c = \frac{3^{1/4} \Gamma(1/4)^2G_1}{8 \sqrt{2 \pi}}\]
along with
\[ \alpha_3 = \alpha_1 = \frac{12^{1/4}\pi}{|2G_1|^{1/3}\vareps^{1/3}}\ee^{2/3G_1\vareps}, \qquad
  c_1 = 12^{1/4}\ee^{2/3G_1\vareps}.\]
Note that there is an exponentially decaying solution reflected back
from the turning point, as there was in a previous work on the nonlinear
Schr\"odinger (NLS) equation~\cite{jon1}, but that the corresponding
contribution to the normal form is an exponentially small correction to an
algebraic series this time, rather than being the dominant term.

In summary, for $|y| \ra \infty$ and $G<0$, we have:
\begin{align*}
w & \sim  12^{1/4}\ee^{-2/3|G_1|\vareps}\frac{1}{(|G_1|y-1)^{1/2}} \\
& =
12^{1/4}\ee^{-2/3|G|}\frac{1}{(|G|\xi-1)^{1/2}}
\sim\frac{ 12^{1/4}\ee^{-2/3|G|}}{|G|^{1/2}\xi^{1/2}} \mbox{ as } \xi \ra \infty,\\
w & \sim \frac{3^{1/4} \Gamma(1/4)^2 |G_1| \vareps}{4 \sqrt{2 \pi}}\frac{1}{(1-|G_1| y )^{1/2}}
\\
& =  \frac{3^{1/4} \Gamma(1/4)^2 |G|}{4 \sqrt{2 \pi}}\frac{1}{(1-|G| \xi )^{1/2}}
\sim
\frac{3^{1/4} \Gamma(1/4)^2}{4 \sqrt{2 \pi}}\frac{|G|^{1/2}}{(-\xi)^{1/2}} %
\mbox{ as } \xi \ra \infty,
\end{align*}
which agrees with~\cite{weinmueller_2020} after correcting the missing
factor of 2 in their expression for $C$.

\subsubsection{$G_1>0$}
For $y>0$ we need $a_3=0$ in Eq.~\eqref{eq:WKB_w_sol} otherwise
we have exponential growth. This way, we have:
\[ w \sim  \frac{a_1}{(1+G_1y)^{1/2}}
  + \frac{a_2}{(1+G_1y)^{1/2}}\ee^{-\frac{2}{3G_1 \vareps}((1+G_1y)^{3/2}-1)},
\]
and in terms of the inner variable
\[ w \sim  \frac{a_1}{(1+G_1 \vareps \xi)^{1/2}}
  + \frac{a_2}{(1+G_1\vareps
    \xi)^{1/2}}\ee^{-\frac{2}{3G_1\vareps}((1+G_1 \vareps\xi)^{3/2}-1)}
\sim a_1 + a_2 \ee^{-\xi} + \cdots.
\]
Matching now with Eq.~\eqref{lead_in}, Eq.~\eqref{in_inf} gives
\beqas
a_1 & = &  \left(-c + \frac{3^{1/4}\Gamma(1/4)^2 G_1}{8 \sqrt{2 \pi}}\right)\vareps,\\
a_2 &=&12^{1/4}.
\eeqas
For $-1/G_1<y<0$ we have
\[ w = \frac{1}{(1+G_1 y)^{1/2}}
\left(b_1+b_2\ee^{-\frac{2}{3G_1\vareps}((1+G_1 y)^{3/2}-1)}%
+ b_3\ee^{\frac{2}{3G_1\vareps}((1+G_1y)^{3/2}-1)}\right),
\]
while for $y<-1/G_1$ we have
\[ w = \frac{1}{(-G_1y-1)^{1/2}}\left(c_1 %
+ c_2\ee^{-\frac{2\ii}{3G_1\vareps}(-G_1y-1)^{3/2}}+
  c_3\ee^{\frac{2\ii}{3G_1\vareps}(-G_1y-1)^{3/2}}\right).
\]
The behaviour at infinity means that $c_2=c_3=0$. Again,
 matching with Eq.~\eqref{lead_in}, Eq.~\eqref{in_minf}
gives $b_2=0$, and
\[ b_3 = 12^{1/4}, \qquad %
b_1 = \left( -c - \frac{3^{1/4}\Gamma(1/4)^2 G_1}{8 \sqrt{2 \pi}}\right)\vareps.
\]
This time, the inner limit of the outer near the turning point is
\beqas
w &=& \frac{b_1}{(2G_1)^{1/3}\vareps^{1/3}z^{1/2}} + \frac{b_3}{(2G_1)^{1/3}z^{1/2}\vareps^{1/3}}\ee^{-2/3G_1\vareps}\ee^{4/3 z^{3/2}} \qquad z>0,\\
w &=& \frac{c_1}{(2G_1)^{1/3}(-z)^{1/2}\vareps^{1/3}}  \qquad z<0.
\eeqas
To match as $z \ra -\infty$ requires
\[\alpha_1=\alpha_3=\frac{c_1\pi }{(2G_1)^{1/3}\vareps^{1/3}}, \qquad \alpha_2=0,\]
whereas as $z \ra \infty$ requires
\[ \alpha_3 = \frac{b_3\pi}{(2G_1)^{1/3}\vareps^{1/3}}\ee^{-2/3G_1\vareps}, \qquad b_1 = 0.
\]
Thus we finally see that
\[ c = -\frac{3^{1/4} \Gamma(1/4)^2G_1}{8 \sqrt{2 \pi}},\]
along with
\[ \alpha_3 = \alpha_1 = \frac{12^{1/4}\pi}{(2G_1)^{1/3}\vareps^{1/3}}\ee^{-2/3G_1\vareps}, \qquad
  c_1 = 12^{1/4}\ee^{-2/3G_1\vareps}.\]

\subsubsection{Comparison of normal form against numerics}

Combining the results for $G_1$ positive and negative, we finally
have
\[ c = - C |G_1|, \qquad C = \frac{3^{1/4}\Gamma(1/4)^2}{8 \sqrt{2 \pi}}.\]
Thus the normal form [cf.~Eq.~\eqref{norm}] becomes
\beq
\alpha C \fdd{G_1}{t} = -|G_1| \beta p_1 + \gamma C
G_1^2.\label{normalform}
\eeq
%

\begin{figure}[pt]
\includegraphics[width = 0.5\textwidth]{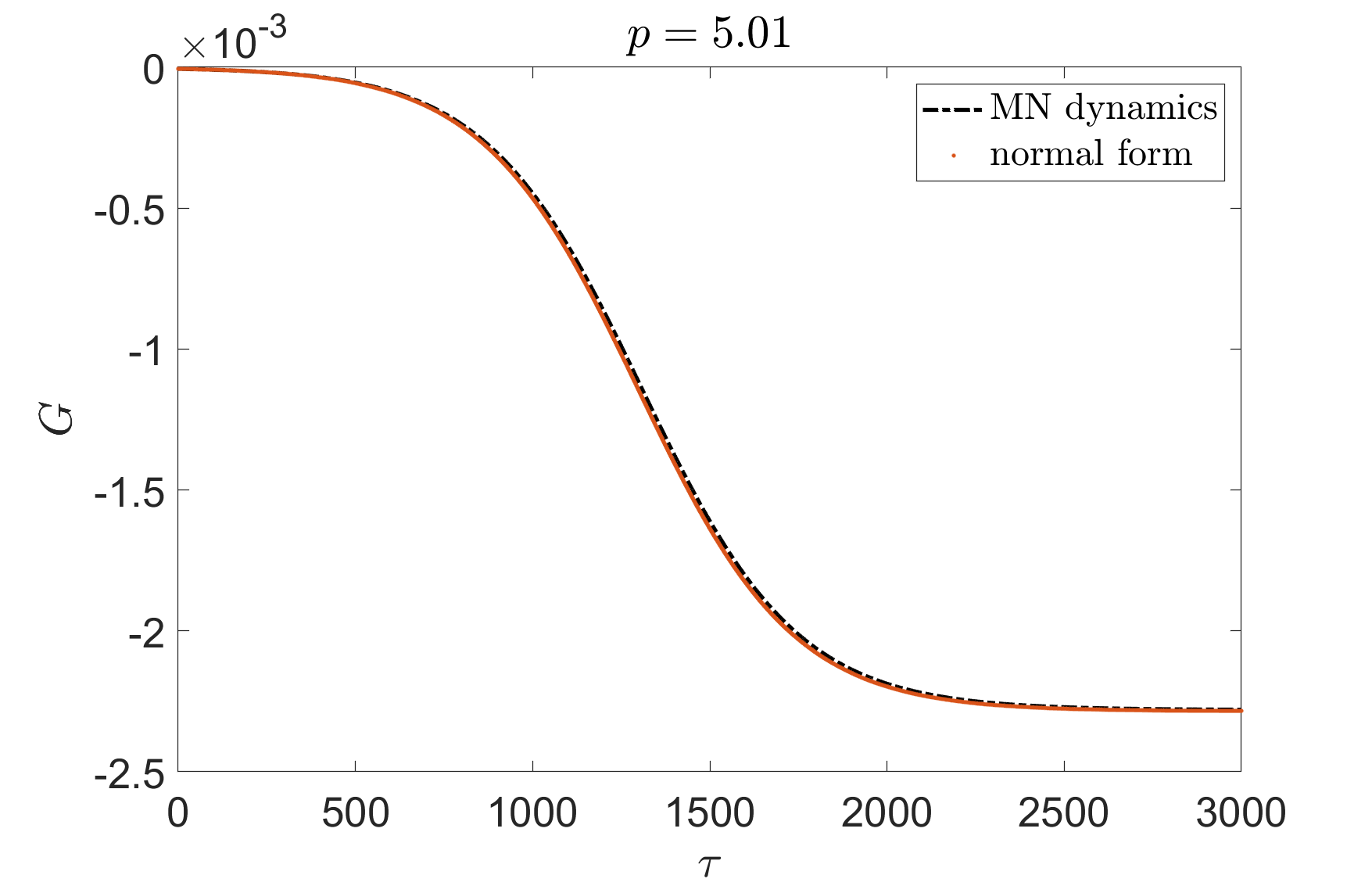}
\includegraphics[width = 0.5\textwidth]{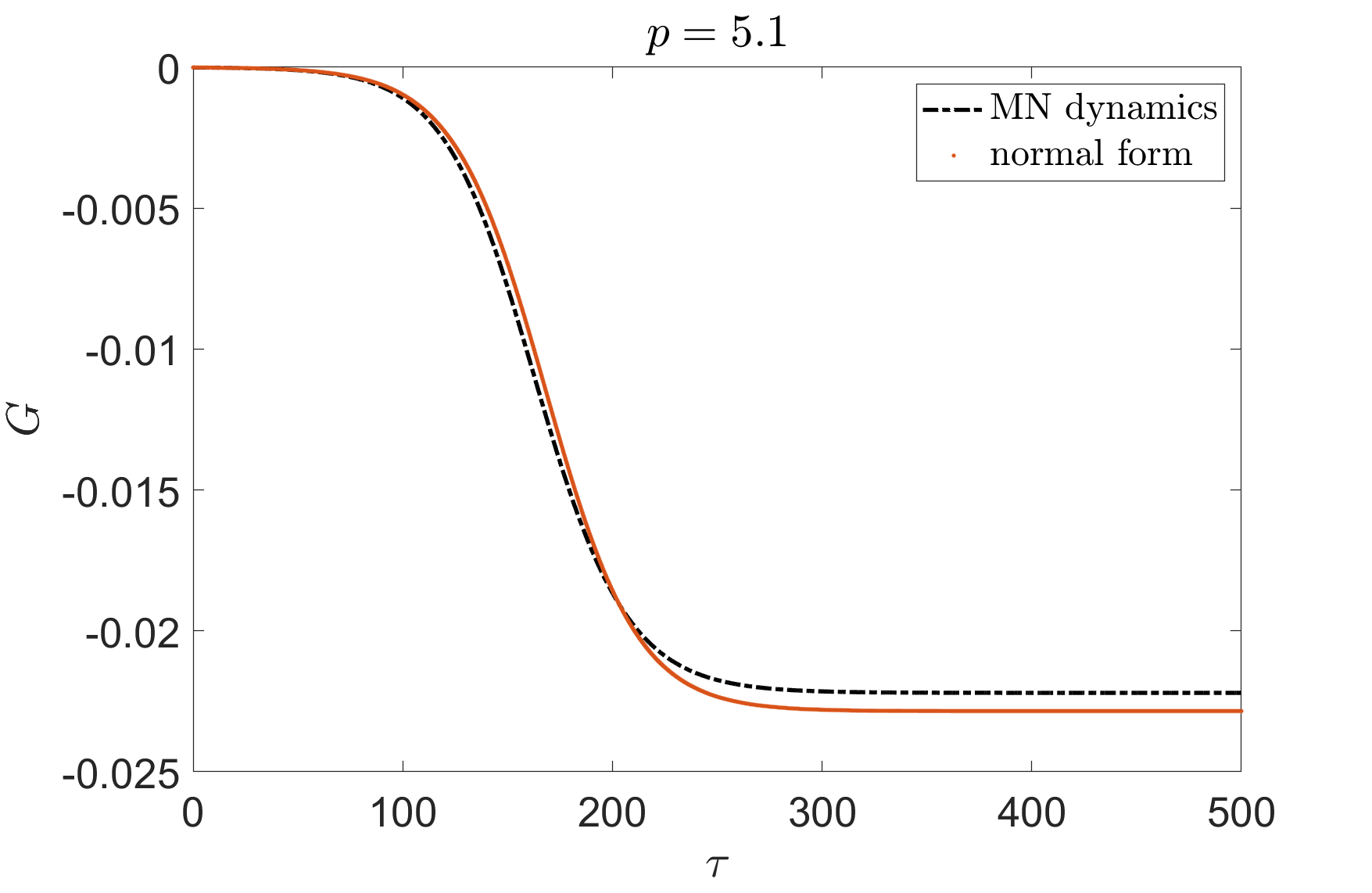}
\caption{The evolution of the blow-up rate $G$ as a function of $\tau$
computed by the self-similar dynamics of Eq.~\eqref{eq:renKdV_v2} and
normal form of Eq.~\eqref{normalform} are shown with a dashed-dotted
black line and red dots, respectively, for values of $p=5.01$ (left panel)
and $p=5.1$ (right panel).
\label{fig:MN_vs_normal501_51}}
\end{figure}

In Fig.~\ref{fig:MN_vs_normal501_51}, we provide a comparative analysis of the blow-up rate computed using the self-similar
dynamics [cf.~Eq.~\eqref{eq:renKdV_v2}] (represented by the dashed-dotted black line) and the normal form (indicated by red dots).
The left panel shows the case for $p=5.01$,
which is in close proximity to the bifurcation point.
Conversely, the right panel illustrates the case for $p=5.1$.
In the left panel, one can observe a very good agreement between the self-similar dynamics and the normal form.
However, as we move away from $p=5$ (as seen in the right panel), a disparity emerges between the two approaches. (see also left panel in Fig.~\ref{fig:blow-up_bifdiagram} which shows the growing disparity between numerics and normal form as we move away from $p=5$)
Notably, both curves still maintain a consistent qualitative trend, indicating a comparable rate of asymptotic convergence towards a self-similar profile.

\section{Conclusions and Future Challenges}

In the present work we have revisited the generalized Korteweg-de Vries
equation model, by considering the bifurcation of its solitary waves
into instability, as the exponent of the generalized nonlinearity
crossed the value of $p=5$. Building on the work of~\cite{koch_2015}
and~\cite{weinmueller_2020}, upon exploring the stability of the
solitary wave solutions with blow-up rate $G=0$, we retrieved the
self-similar solutions with non-vanishing blow-up rate $G \neq 0$
which were found both dynamically, as well as statically as stationary
states in the so-called co-exploding frame. One of the key contributions of
the present work involves the elucidation of the
bifurcation
associated with the emergence of this self-similar branch of solutions
past the critical point of $p=5$; we found this bifurcation
to bear features of a pitchfork (through the presence
of two mirror symmetric blow-up solutions), as we as to
encompass linear (in absolute value) and quadratic terms in
its normal form (a feature somewhat reminiscent of a transcritical
normal form). We have discussed both the similarities
and the differences of the relevant phenomenology from that of the
other prototypical dispersive nonlinear PDE, namely the NLS which features
a similar bifurcation, but an exponentially deviating branch past the
corresponding critical point. The second key contribution of the present
work concerns the spectral stability analysis of the emerging self-similar branch.
In connection with the latter, we elucidate both the trajectory of the eigenvalues of the point spectrum, such as the ones at $-3 G$ (due to
the former scaling invariance in the original frame) or at $-G$ (due to
the former translational invariance at the original frame), or the one
at (nearly) $G/2$, but also those of the continuous spectrum. The latter
consists of two main bands, one of which aligns (as it is better resolved)
with the vertical line $\lambda=G$, while the other we argued to be
extremely ill-conditioned
and cannot be resolved.
If they could be resolved,
these eigenvalues would move to
the negative real axis.
Instead, what we find is that
we can only resolve the first
few of them. Moreover,  the rest becomes harder to resolve as the
domain size $L$ gets larger,  leading the corresponding eigenvalue wedge to become more
vertical rather than more horizontal. This is rather telling as regards the complexity of the corresponding spectrum and the difficulty to
numerically pin it down.

The addressing of the above challenges
paves the way for the further consideration
of a number of interesting problems
in this context of self-similar
dynamics. As concerns the case of
the generalized KdV, having addressed
the existence and stability of the
self-similar states, it would be useful
to understand better the fully nonlinear
dynamics in both the original (non-exploding)
and the presently considered (co-exploding)
frame in line with important earlier
dynamical observations, such as those
of, e.g.,~\cite{klein_peter_2015}.
In particular, the dynamical reshaping
of the structure (for different
initial conditions/moment) into
a self-similarly collapsing waveform
constitutes an intriguing dynamical question.
It is also interesting to extend relevant
existence/stability/dynamical considerations
beyond conservative cases such as the
generalized NLS and KdV, where it seems
that the emergent instabilities in the
co-exploding frame are only apparent/fictitious
ones related to symmetries of the original
frame, but not true dynamical instabilities.
In that vein, a particularly intriguing
example to consider concerns the
non-conservative case of the complex
Ginzburg-Landau equation
earlier considered in the work of~\cite{budd_05}. There, there exist
both branches that stem from the NLS
and once that do not (in the limit of
small dissipation) and both branches
that are stable and ones that
are unstable in the co-exploding frame.
Hence, it will be quite interesting
to expand existence/stability, but
also importantly dynamics considerations
to such states in future works.

\section*{Acknowledgements}
EGC expresses his gratitude to Professors Uri~M.~Ascher (University of
British Columbia), Raymond Spiteri (University of Saskatchewan),
Eva Weinm\"uller (Vienna University of Technology) as well as Linda~Petzold
(University of California Santa Barbara) for fruitful discussions.
This work has been supported by the U.S. National Science Foundation
under Grants DMS-2204782 (EGC) and DMS-2110030 and DMS-2204702 (PGK).
The work of IGK was partially supported by the
US-DoE.





\begin{thebibliography}{00}
\bibitem{jon_expl_1}
Note that the work of~\cite{weinmueller_2020} does not include the
first term on the RHS of Eq.~\ref{w1eqn}. However, omitting this
term does not affect the solvability condition. Essentially the
analysis in~\cite{weinmueller_2020} perturbs only the $p$ which
appears as a coefficient, not that which appears as an exponent,
that is, they solve
%
\begin{align*}
\pd{w}{\tau} &= - \frac{\dd^3 w}{\dd\xi^3} %
- \pd{w^5}{\xi} + G\left(\frac{2}{p-1} w + \xi \pd{w}{\xi} \right) + \pd{w}{\xi}.
\end{align*}
\end{thebibliography}


\bibliographystyle{elsarticle-harv}

\end{document}